\documentclass[useAMS,usenatbib]{mn2e}
\usepackage{epsfig}

\newcommand{\mwfrac}[1]{{\langle x_{\rm #1}}\rangle_{\rm M}}

\newcommand{\mHt}{{\rm H_{2}}}
\newcommand{\me}{{\rm e^{-}}}
\newcommand{\mH}{{\rm H}}
\newcommand{\Hp}{\rm{H}^{+}}
\newcommand{\mHtp}{\rm{H}_{2}^{+}}
\newcommand{\htp}{\rm{H}_{3}^{+}}
\newcommand{\He}{\rm{He}}
\newcommand{\Hep}{\rm{He}^{+}}
\newcommand{\Cp}{\rm{C^{+}}}
\newcommand{\mC}{\rm{C}}
\newcommand{\chx}{{\rm CH_{x}}}
\newcommand{\mO}{\rm{O}}
\newcommand{\ohx}{\rm{OH_{x}}}
\newcommand{\co}{{\rm CO}}
\newcommand{\hcop}{{\rm HCO^{+}}}
\def\simless{\mathbin{\lower 3pt\hbox
   {$\rlap{\raise 5pt\hbox{$\char'074$}}\mathchar"7218$}}}
\def\simgreat{\mathbin{\lower 3pt\hbox
   {$\rlap{\raise 5pt\hbox{$\char'076$}}\mathchar"7218$}}}

\title[Approximations for CO chemistry]{Approximations for modelling CO chemistry in GMCs: a comparison of approaches}
\author[Glover \& Clark]{Simon~C.~O.~Glover \& Paul~C.~Clark \\
Zentrum f\"ur Astronomie der Universit\"at Heidelberg, Institut f\"ur Theoretische
Astrophysik, Albert-Ueberle-Str.\ 2, 69120 Heidelberg \\
 {\tt email:} glover@uni-heidelberg.de, p.clark@uni-heidelberg.de}

\begin{document}

\maketitle

\begin{abstract}
 We examine several different simplified approaches for modelling the 
 chemistry of CO in three-dimensional numerical simulations of turbulent
 molecular clouds. We compare the different models both by looking at
 the behaviour of integrated quantities such as the mean CO fraction 
 or the cloud-averaged CO-to-H$_{2}$ conversion factor, and also by
 studying the detailed distribution of CO as a function of gas density
 and visual extinction. In addition, we examine the extent to which the
 density and temperature distributions depend on our choice of chemical
 model.
 
 We find that all of the models predict the same density PDF and also
 agree very well on the form of the temperature PDF for temperatures
 $T > 30 \; {\rm K}$, although at lower temperatures, some difference
 become apparent. All of the models also predict the same CO-to-H$_{2}$ 
 conversion factor, to within a factor of a few. However, when we look
 more closely at the details of the CO distribution, we find larger 
 differences. The more complex models tend to produce less CO and
 more atomic carbon than the simpler models, suggesting that the C/CO
 ratio may be a useful observational tool for determining which model
 best fits the observational data. Nevertheless, the fact that these chemical
 differences do not appear to have a strong effect on the density or temperature
distributions of the gas suggests that the dynamical behaviour of the molecular 
clouds on large scales is not particularly sensitive to how accurately the 
small-scale chemistry is modelled.
\end{abstract}

\begin{keywords}
galaxies: ISM -- ISM: clouds -- ISM: molecules -- molecular processes
\end{keywords}

\section{Introduction}
Carbon monoxide (CO) is a key constituent of the gas making up the giant molecular
clouds (GMCs) that are the site of almost all Galactic star formation. Although almost all of the
molecular gas mass within a GMC is in the form of molecular hydrogen (H$_{2}$), this
material is extremely difficult to observe directly, as the temperatures within a GMC are
too low to excite even the lowest accessible rotational transition of the H$_{2}$ molecule,
the quadrupole transition between the $J=2$ and $J=0$ rotational levels. On the
other hand, CO does become rotationally excited at GMC temperatures, owing to the
much smaller energy separation between its rotational levels. It is readily observed in the
millimeter, and is arguably the single most important observational tracer 
of the state of the gas within the clouds. It also plays an important role as the main
molecular coolant of the gas over a wide range in densities \citep[see e.g.][]{nlm95}. 
It is therefore important to understand the distribution of CO within GMCs, and how this relates to
the underlying gas distribution.

Unfortunately, this is not a simple task. It has been understood for a long time that the
gas within GMCs is typically not in chemical equilibrium \citep[see e.g.][]{lhh84}. More recently,
numerical modelling has highlighted the fact that the CO abundance within a given
parcel of gas within a GMC is a complex function of the density and temperature of the
gas, the local radiation field (modulated by absorption by gas elsewhere in the cloud),
and possibly also the dynamical history of the gas \citep{glo10}. It has also become 
increasingly clear that molecular clouds are dynamically complex objects, dominated
by supersonic, turbulent motions \citep[see e.g.][for a comprehensive overview]{mk04},
and that we ignore the effects of this dynamical complexity at our peril. Therefore, if we
want to be able to accurately model the formation and destruction of CO within a GMC,
the use of detailed models that treat both the chemistry and the turbulent dynamics in
an appropriately coupled fashion seems unavoidable.

However, this presents us with a serious numerical challenge. Even if we focus only 
on the chemistry of hydrogen, helium, carbon and oxygen, and ignore other elements such 
as nitrogen or sulphur, there are still a very large number of possible reactions and
reactants that we could include in our chemical model. For example, in the 2006 release
of the UMIST Database for Astrochemistry \citep{umist07}, there are 2115 such reactions, involving 
a total of 172 different 
chemical species. In particular, it is the large number of chemical species that must
potentially be handled that is the main source of our problems. If we model the chemical
evolution of the gas by using a set of ordinary differential equations (ODEs) to represent
the rates of change of the chemical abundances, then we find that the required set of
ODEs  is generally extremely stiff, containing processes with a wide range of different
intrinsic timescales. To evolve this set of ODEs in a numerically stable fashion, we must
therefore either evolve them explicitly in time using extremely small timesteps, which is
impractical, or evolve them implicitly in time. However, if we adopt an implicit approach,
then we find in general that the computational cost of solving the implicit ODEs scales as 
the cube of the number of species involved, $N_{\rm spec}^{3}$. Therefore, when 
$N_{\rm spec}$ is large, the cost of solving the chemical rate equations can easily come
to dominate the total computational cost, putting strict limits on the size of the problems
that can be tackled.  In practice, values of $N_{\rm spec}$ as large as 14 can be handled
in three-dimensional turbulence simulations at the present day (see e.g.\ \citealt{glo10};
note that while their model includes 32 distinct chemical species, all but 14 of these 
are assumed to be in chemical equilibrium, or are trivially derivable from conservation laws), 
but even in this case, solving the chemical rate 
equations can take up as much as  90\% of the total runtime of the simulation. Scaling up 
from here to the $N_{\rm spec} =172$ subset of the UMIST database mentioned above
remains computationally impractical in three-dimensional simulations.

For this reason, a number of authors have sought to reduce $N_{\rm spec}$ to a more
tractable value by identifying and retaining only the most important of the reactions 
involved in the formation and destruction of CO, as well as making other simplifications
that are discussed in later sections. When identifying the key reactions, there is an obvious 
trade-off between complexity (and hence computational cost) and completeness: the
larger the number of possible chemical pathways that we include, the slower the code will
run. There are models available in the literature that span a range of different complexities, 
but until now there has been no comparison of the results produced by these different
approaches.

In this paper, we present results from a detailed comparison of several different approximate
methods for modelling CO formation and destruction in turbulent molecular clouds that span
a range of different complexities. We are interested in particular in establishing which results 
from simulations of turbulent clouds are highly sensitive to the choice of simplified chemical
model, and which are relatively insensitive.

\section{Simulation setup}
\label{setup}
The basic setup of the simulations used here is similar to that described in \citet{glo10}
and \citet{gm10}. We use the ZEUS-MP code to  
model driven magnetohydrodynamical turbulence in a cubical 
volume, with side length $L = 20 \: {\rm pc}$, using periodic boundary conditions for the 
gas. The turbulence is driven using the algorithm described in \citet{mkbs98}, and the amplitude 
of  the driving is chosen in order to maintain the rms turbulent velocity at  $v_{\rm rms} = 
5 \: {\rm km} \: {\rm s^{-1}}$. The initial magnetic field strength is $B_{0} = 5.85 \: \mu{\rm G}$, 
and the field is assumed to be initially uniform and oriented parallel to the $z$-axis of the simulation.
The effects of self-gravity are not included. The simulations are run for $t = 1.8 \times 10^{14} \:
{\rm s} \simeq 5.7 \: {\rm Myr}$, or approximately three turbulent eddy turnover times. We showed
in \citet{glo10} that this is long enough to allow the CO distribution to settle into a statistical steady
state.

We perform two sets of simulations (hereafter denoted as set 1 and set 2) with 
different initial densities: $n_{0} = 100 \: {\rm cm^{-3}}$ for set 1 and $n_{\rm 0}
= 300 \: {\rm cm^{-3}}$ for set 2, where $n_{0}$ is the initial number density of
hydrogen nuclei. The corresponding mean extinctions are therefore $\bar{A}_{\rm V}
= 3.3$ and $\bar{A}_{\rm V} = 9.9$, respectively, where we have assumed a conversion 
factor from hydrogen column density to dust visual extinction of 
$5.348 \times 10^{-22} \mbox{ mag cm}^{2}$, appropriate to dust in the cold interstellar 
medium (ISM). We therefore expect that in both cases, detectable amounts of CO will be 
produced in the clouds, even though the mean mass-weighted CO abundance
should differ by more than an order of magnitude between the two sets of runs 
\citep{gm10}.

 In both sets of simulations, we adopt elemental abundances
(by number, relative to hydrogen) of $x_{\rm He} = 0.1$, $x_{\rm C} = 1.41
\times 10^{-4}$ and $x_{\rm O} = 3.16 \times 10^{-4}$ for helium, carbon and
oxygen, respectively. We assume that the hydrogen, helium and oxygen are initially
atomic, and that the carbon is initially in singly ionized form throughout the simulation
volume.

We set the initial gas temperature to 60~K and the initial dust temperature to 10~K
in both sets of simulations. To model the subsequent heating and cooling of the 
gas and the dust, we use a modified version of the \citet{glo10} thermal model,
as described in that paper and in the Appendix. Set 1 and set 2 consist of six 
simulations each,  one for each of the chemical models described in 
Section~\ref{chem_model}. In the models that include the effects of cosmic rays,
we adopt a rate $\zeta_{\rm H} =  10^{-17} \: {\rm s^{-1}}$ for the cosmic ray ionization of atomic 
hydrogen. Finally, to model the effects of the external ultraviolet 
radiation field, we assume that its strength and shape are the same as those of the
standard \citet{dr78} field. We illuminate each side of the simulation box using the
unattenuated field, and model attenuation within the box due to dust absorption,
H$_{2}$ self-shielding, CO self-shielding and the shielding of CO by H$_{2}$ using
the `six-ray' approximation introduced by \citet{nl97}. Full details of this procedure
can be found in \citet{glo10}.

\section{Chemical models}
\label{chem_model}
\subsection{Glover et~al.~(2010) [G10g, G10ng]}
\label{ref}
The first model we consider is a slightly modified version of the \citet{glo10} chemical 
model. This consists of 218 reactions amongst 32 species, and so although it is the
most complicated of the approximate models included in this study, it nevertheless
still represents a considerable simplification compared to a model including
the full set of reactions from e.g.\ the UMIST Database for Astrochemistry. However,
\citet{glo10} demonstrated that this simplified model could accurately reproduce the 
results obtained with a far more comprehensive model for several 1D test problems,
while at the same time being small enough to be usable in a three-dimensional 
simulation. 

We have made one significant change to the \citet{glo10} chemical network: the inclusion
of an optional treatment of the effects of the recombination of H$^{+}$, He$^{+}$, C$^{+}$, 
and O$^{+}$ ions on the surfaces of charged dust grains. We treat these processes using
the formalism of \citet{wd01}, which includes the effects of very small dust grains and
polycyclic aromatic hydrocarbons. These grain surface processes are not included in any
of the other models that we examine, and in order to understand what effect they have on
the outcome of the simulations, we have performed runs both with and without them. In
the remainder of the paper, we denote these simulations as G10g and G10ng, respectively.

We have also made a number of modifications to our treatment of the thermodynamic
behaviour of the gas, in order to improve our ability to model very cold gas. These 
improvements are used for all of the models examined here and full details of them are 
given in the Appendix.

\subsection{Nelson \& Langer (1997) [NL97]}
A much simpler approximation is given by the model proposed by \citet{nl97},
which they used
to study the dynamics of low-mass ($M = 100$--$400 \: {\rm M_{\odot}}$) molecular clouds. In
their study, \citet{nl97} assume that all of the hydrogen is already in the form of H$_{2}$, and
focus their attention on the conversion of singly ionized carbon, C$^{+}$, to carbon monoxide, CO.
Their approximation involves the assumption of direct conversion from C$^{+}$ to CO, and vice
versa, which allows them to ignore any intermediate species (such as neutral atomic carbon, C).
They assume that the conversion of C$^{+}$ to CO is initiated by the formation of an intermediate
hydrocarbon radical (e.g.\ CH or CH$_{2}$), which they denote as CH$_{x}$. This may then react
with oxygen to form CO, or be photodissociated by the interstellar radiation field. Once CO has 
formed, it is then only destroyed by photodissociation, yielding C and O, but the neutral
carbon produced in this way is assumed to be instantly photoionized, yielding C$^{+}$. Since 
the formation of the hydrocarbon radical will typically involve a slow radiative association reaction 
as the initial step, such as the formation of CH$_{2}^{+}$ via
\begin{equation}
{\rm C}^{+} + {\rm H_{2}} \rightarrow {\rm CH}_{2}^{+} + \gamma,
\end{equation}
\citet{nl97} assume that this is the rate limiting step for the formation of CO, and write the rate
equation for the CO number density as:\footnote{In actual fact, Nelson \& Langer give slightly different forms for Equations~\ref{nl97_1} and \ref{nl97_2}, writing $n$ in place of $n_{\rm H_{2}}$ in 
Equation~\ref{nl97_1} (equation~18 in their paper) and vice versa in Equation~\ref{nl97_2}
(equation 20 in their paper). However, this appears to be a
typographical error, as one can see by considering the behaviour of the system when the 
H$_{2}$ number density and the UV field strength are both zero. According to the logic of the
NL97 model, the formation rate of CO in this case should also be zero, since H$_{2}$ is required 
in order to form the intermediate hydrocarbon radical. However, if one uses the original expressions
given in \citet{nl97} for $\frac{{\rm d}n_{\rm CO}}{{\rm d}t}$ and $\beta$, one finds that they do not
show this behaviour -- instead, $\beta \rightarrow 1$ in the limit that $n_{\rm H_{2}} \rightarrow 0$
and the predicted CO formation rate remains larger than zero.  }
 \begin{equation}
 \label{nl97_1}
\frac{{\rm d}n_{\rm CO}}{{\rm d}t} = k_{0} n_{\rm C^{+}} n_{\rm H_{2}} \beta - 
\Gamma_{\rm CO} n_{\rm CO}, 
\end{equation}
where $n_{\rm C^{+}}$ is the number density of C$^{+}$ ions and $n_{\rm H_{2}}$ is the number
density of hydrogen molecules. In Equation~\ref{nl97_1}, $k_{0}$ is the rate coefficient for the formation of the 
intermediate CH$_{x}$ ion or radical,
which \citet{nl97} give as $k_{0} = 5 \times 10^{-16} \: {\rm cm^{3}} \: {\rm s^{-1}}$, and $\Gamma_{\rm CO}$
is the photodissociation rate of CO, given in their model as
\begin{equation}
\Gamma_{\rm CO} = 10^{-10} G_{0} \exp(-2.5 A_{\rm V}) \: {\rm s^{-1}},
\end{equation}
where $G_{0}$ is the strength of the ultraviolet radiation field in units of the \citet{habing68} field. In
our simulations, we adopt the \citet{dr78} parameterization of the interstellar ultraviolet radiation field
and so $G_{0} = 1.7$. The variable $\beta$ in Equation~\ref{nl97_1} represents the proportion of the CH$_{x}$ that 
successfully forms CO. This is given in the \citet{nl97} model by
\begin{equation}
 \label{nl97_2}
\beta = \frac{k_{1} x_{\rm O}}{k_{1} x_{\rm O}  + \Gamma_{\rm CH_{x}} / n},
\end{equation}
where $n$ is the number density of hydrogen nuclei,
$x_{\rm O}$ is the fractional abundance of atomic oxygen, $k_{1}$ is the rate coefficient for
the formation of CO from O + CH$_{x}$, given by \citeauthor{nl97} as $k_{1} = 5 \times 10^{-10}
\: {\rm cm^{3}} \: {\rm s^{-1}}$,
and $\Gamma_{\rm CH_{x}}$ is the photodissociation rate of CH$_{x}$, given by
\begin{equation}
\Gamma_{\rm CH_{x}} = 5 \times 10^{-10} G_{0} \exp(-2.5 A_{\rm V}) \: {\rm s^{-1}}.
\end{equation}

We have implemented this treatment of the carbon chemistry, with a couple of minor changes. In place of the rate assumed by \citet{nl97} for $\Gamma_{\rm CO}$, 
we use a rate $\Gamma_{\rm CO} = 2 \times 10^{-10} G_{0} \exp(-2.5 A_{\rm V}) f_{\rm sh}  \: {\rm s^{-1}}$, where $f_{\rm sh}$ is a shielding factor
that quantifies the effects of CO self-shielding and the shielding of CO by H$_{2}$ Lyman-Werner band
absorption. This is the same rate coefficient as that used in the G10g model, and is based on work by \citet{vdb88} and \citet{lee96}. We have made this substitution in an effort to minimize any differences between the models that arise purely from differences in the rate coefficients adopted. The influence of rate coefficient uncertainties on molecular cloud chemistry has been studied in detail elsewhere 
\citep[see e.g.][]{mil88,wak06,wak10} and is not our primary focus here, as we are interested more in 
the influence of the design of the chemical network itself. Since we have adopted a larger value for 
$\Gamma_{\rm CO}$, we have also adopted a larger value for $\Gamma_{\rm CH_{x}}$, so as to
keep the ratio of $\Gamma_{\rm CH_{x}} / \Gamma_{\rm CO}$ in the optically thin limit 
the same as in \citet{nl97}. This yields a value for $\Gamma_{\rm CH_{x}}$ that is more in keeping 
with the rates adopted for CH and CH$_{2}$ photodissociation and photoionization in the G10g model \citep[see][Appendix A]{glo10}. To compute the shielding factor $f_{\rm sh}$, we use our standard
six-ray treatment to determine the H$_{2}$ and CO column densities, and then convert these to a
shielding factor using the data tabulated in \citet{lee96}.

In addition, and unlike \citet{nl97}, we do not assume that the hydrogen is completely molecular, but
instead follow the evolution of the H$_{2}$ and H$^{+}$ abundances explicitly using the same 
hydrogen chemistry network as in \citet{gm07a,gm07b}. The abundance of neutral atomic hydrogen,
H, then follows from a simple conservation law. We include the effects of H$_{2}$ self-shielding and
dust shielding using the same six-ray treatment as in the G10g model.

\subsection{Nelson \& Langer (1999) [NL99]}
In a later paper, Nelson \& Langer suggested an alternative approximation for modelling the formation
of CO \citep{nl99}. This was designed for a
similar purpose as the \citet{nl97} approximation, but is considerably more sophisticated. Notably,
it allows for the formation of CO via multiple pathways. In addition to the formation channel involving
the composite hydrocarbon radical CH$_{x}$ (which should be understood to represent both CH and 
CH$_{2}$) that forms the basis of the NL97 model, the \citet{nl99} model (hereafter NL99) also allows for CO
formation via the composite oxygen species OH$_{x}$ (representing the species OH, H$_{2}$O, 
O$_{2}$ and their ions)  as well as via the recombination of HCO$^{+}$.
In addition, photodissociation is no longer the only fate for the CO: the network also includes  the
conversion of CO to HCO$^{+}$ by proton transfer from H$_{3}^{+}$, and its destruction by dissociative
charge transfer from ionized helium:
\begin{equation}
{\rm CO} + {\rm He^{+}} \rightarrow {\rm C^{+}} + {\rm O} + {\rm He}.
\end{equation}
A further notable difference between the NL97  and NL99 models is the fact that the latter model tracks
the abundance of neutral atomic carbon, rather than just C$^{+}$ and CO.
Finally, \citet{nl99} also include in their model a small number of reactions involving a species they denote as M
that represents the combined effects of low ionization potential metals such as Mg, Fe,
Ca and Na, which become the dominant atomic charge carriers in very shielded regions of
the cloud. The full list of reactions included in the \citet{nl99} model is given in Table~\ref{tab:nl99}.

\begin{table}
\caption{Reactions in the NL99 chemical model \label{tab:nl99}}
\begin{tabular}{lc}
Reaction & Notes \\
\hline
$\mHt + {\rm c.r.} \rightarrow \mHtp + \me$ & 1 \\
$\mHtp  + \mHt \rightarrow \htp + \mH$ & 1 \\
${\rm He} + {\rm c.r.} \rightarrow \Hep + \me$ & \\
$\mC + \htp \rightarrow \chx + \mHt$ & \\
$\mO + \htp \rightarrow \ohx + \mHt$ & \\
$\co + \htp \rightarrow \hcop + \mHt$ & \\
$\Hep + \mHt \rightarrow \He + \mH + \Hp$ & \\
$\Hep + \co \rightarrow \Cp + \mO + \He$ & \\
$\Cp + \mHt \rightarrow \chx + \mH$ & \\
$\Cp + \ohx \rightarrow \hcop$ & \\
$\mO + \chx \rightarrow \co + \mH$ & \\
$\mC + \ohx \rightarrow \co + \mH$ & \\
$\Hep + \me \rightarrow \He + \gamma$ & \\
$\htp + \me \rightarrow \mHt  + \mH$ & \\
$\Cp + \me \rightarrow \mC + \gamma$ & \\
$\hcop + \me \rightarrow \co + \mH$ & \\
${\rm M^{+}} + \me \rightarrow {\rm M} + \gamma$ & 2 \\
$\htp + {\rm M} \rightarrow {\rm M^{+}} + \mHt + \mH$ & 2 \\
$\mC + \gamma \rightarrow \Cp + \me$ & \\
$\chx + \gamma \rightarrow \mC + \mH$ & \\ 
$\co  + \gamma \rightarrow  \mC + \mO$ & \\
$\ohx + \gamma \rightarrow  \mO + \mH$ & \\
${\rm M} + \gamma \rightarrow {\rm M^{+}} + \me$ & 2 \\
${\rm HCO^{+}} + \gamma \rightarrow \co + \Hp$ & \\
\hline
\end{tabular}
\medskip
\\
{\bf Notes:}
{\bf 1:} These two reactions are combined into a single pseudo-reaction in NL99, as it is assumed that
all of the $\mHtp$ formed by the first reaction is immediately consumed by the second.
{\bf 2:} M represents the combined contributions of several low ionization potential metals, such
as  Mg, Fe, Ca and Na.  
\end{table}

Many of the reactions in the NL99 model  are also included in the G10g model,
and for these reactions we adopt the same rate coefficients in both models, for the reasons discussed 
previously. For the reactions not in the G10g model -- notably, those involving 
CH$_{x}$, OH$_{x}$ or M, we adopt the same rate coefficients as in NL99. For the elemental
abundance of M, which is not included in any of our other chemical models, we adopt the same
value as in \citet{nl99}, i.e.\ $x_{\rm M, tot} = 10^{-7}$, and we assume that it is initially fully
ionized.

We also supplement the list of reactions given in Table~\ref{tab:nl99} with those used in the 
\citet{gm07a,gm07b} network for hydrogen chemistry, and once again include the effects of  
shielding using our standard six-ray approach. As in the case of the NL97 model, we include
the effects of CO self-shielding and the shielding of CO by H$_{2}$ in order to allow us to make
a fair comparison with the G10g model.

\subsection{Keto \& Caselli (2008) [KC08e, KC08n]}
The final two approximations that we consider in this study are based on the work of
\citet{kc08}, and were developed for the study of the thermal balance in dense prestellar
cores.  As in \citet{nl97}, they assume that CO forms primarily via an intermediate 
hydrocarbon radical, explicitly assumed in this case to be CH$_{2}$, and that the formation
of this radical is the rate limiting step in the formation of CO. Unlike \citet{nl97}, they do not
account for photodissociation of the CH$_{2}$, and so write the rate equation for the CO
number density as
\begin{equation}
\frac{{\rm d}n_{\rm CO}}{{\rm d}t} = k_{\rm RA} n_{\rm H_{2}} n_{\rm C^{+}} - \Gamma_{\rm CO} 
n_{\rm CO}, \label{kc-eq1}
\end{equation}
where $k_{\rm RA}$ is the rate coefficient describing the formation of CH$_{2}$ from C$^{+}$
and H$_{2}$ (via radiative association to form CH$_{2}^{+}$, which is then rapidly converted
to CH$_{2}$) and $\Gamma_{\rm CO}$ is the photodissociation rate of CO, the only destruction mechanism for CO that is included in their models. Unlike \citet{nl97}, they do not assume that 
the carbon produced by CO photodissociation will be instantly photoionized, and hence write the 
rate equation for the neutral carbon number density as
\begin{equation}
\frac{{\rm d}n_{\rm C}}{{\rm d}t} =  \Gamma_{\rm CO} n_{\rm CO} - \Gamma_{\rm C} n_{\rm C},
\label{kc-eq2}
\end{equation}
where  $\Gamma_{\rm C}$ is the photoionization rate of atomic carbon. The rate equation for the 
C$^{+}$ number density then follows as
\begin{equation}
\frac{{\rm d}n_{\rm C^{+}}}{{\rm d}t} = \Gamma_{\rm C} n_{\rm C} -  k_{\rm RA} n_{\rm H_{2}} 
n_{\rm C^{+}}. \label{kc-eq3}
\end{equation}
In their study, \citet{kc08} assume chemical equilibrium, and hence eliminate all of 
the time derivatives from the above set of equations, yielding a set of coupled 
algebraic equations. If one also makes use of the conservation equation relating the
fractional abundance of C$^{+}$, C and CO to the total fractional abundance of carbon,
$x_{\rm C, tot}$, 
\begin{equation}
x_{\rm C^{+}} + x_{\rm C} + x_{\rm CO} = x_{\rm C, tot},
\end{equation}
then these algebraic equations are simple to solve for the equilibrium fractional 
abundances of C$^{+}$, C and CO. One obtains:
\begin{equation}
\frac{x_{\rm C^{+}}}{x_{\rm C, tot}} =
\left( 1 + \frac{x_{\rm CO}}{x_{\rm C^{+}}} + \frac{x_{\rm C}}{x_{\rm C^{+}}} \right)^{-1},
\end{equation}
\begin{equation}
\frac{x_{\rm C}}{x_{\rm C, tot}} = \left(\frac{x_{\rm C}}{x_{\rm C^{+}}} \right)
\left( 1 + \frac{x_{\rm CO}}{x_{\rm C^{+}}} + \frac{x_{\rm C}}{x_{\rm C^{+}}} \right)^{-1},
\end{equation}
and
\begin{equation}
\frac{x_{\rm CO}}{x_{\rm C, tot}} = \left(\frac{x_{\rm CO}}{x_{\rm C^{+}}} \right)
\left( 1 + \frac{x_{\rm CO}}{x_{\rm C^{+}}} + \frac{x_{\rm C}}{x_{\rm C^{+}}} \right)^{-1},
\end{equation}
where the ratio of CO to C$^{+}$ is given by\footnote{Note that Equation~5 in \citet{kc08}, which
gives $x_{\rm CO} / x_{\rm C^{+}}$ as $\Gamma_{\rm CO} / \Gamma_{\rm C}$ is incorrect. The 
authors have kindly confirmed to us that this is due to a typographical error: $\Gamma_{\rm CO} / \Gamma_{\rm C}$ actually corresponds to the ratio of the C and CO abundances, $x_{\rm C} / x_{\rm CO}$, rather than $x_{\rm CO} / x_{\rm C^{+}}$ as printed. The authors have verified that this error 
only appears in the journal article, and not in the original numerical modeling, and hence it does not 
affect any of the results presented in \citet{kc08}.}
\begin{equation}
\frac{x_{\rm CO}}{x_{\rm C^{+}}} = \frac{k_{\rm RA} n_{\rm H_{2}}}{\Gamma_{\rm CO}},
\end{equation}
and the ratio of C to C$^{+}$ is given by
\begin{equation}
\frac{x_{\rm C}}{x_{\rm C^{+}}} = \frac{k_{\rm RA} n_{\rm H_{2}}}{\Gamma_{\rm C}}.
\end{equation}
\citet{kc08} adopt numerical values for $k_{\rm RA}$, $\Gamma_{\rm CO}$ and
$\Gamma_{\rm C}$ from \citet{th85}, but as in the other models we examine, we
instead use values from \citet{glo10} in order to minimize any differences arising
purely from  differences in the adopted rate coefficients. As in the other models, we
account for the shielding of CO by CO and H$_{2}$ when computing $\Gamma_{\rm CO}$.

We have implemented this equilibrium treatment of the carbon chemistry and coupled it with
our standard treatment of the non-equilibrium hydrogen chemistry, described in \citet{gm07a}.
We denote this implementation as KC08e (for ``equilibrium'') in the remainder of the paper. We 
have also implemented a non-equilibrium carbon chemistry based on the same set of reactions, 
but which uses the time-dependant rate equations  (Eqs.~\ref{kc-eq1}--\ref{kc-eq3}) as its basis, 
rather than the equilibrium abundances. This  implementation is denoted below as KC08n 
(for ``non-equilibrium''); note also that \citet{kc10} use a similar non-equilibrium scheme.

\section{Performance}
\label{perf}
We begin our comparison of these various approaches to modelling CO formation and
destruction in turbulent gas by examining their relative computational performance.
In Table~\ref{tab:perf} we list the computational time required (in units of CPU hours) to perform
the simulations discussed in this paper. All of the simulations were run on 32 cores on {\it kolob},
an Intel Xeon Quad-Core cluster at the University of Heidelberg.\footnote{For full technical details,
see http://kolob.ziti.uni-heidelberg.de/kolob/htdocs/hardware/technical.shtml}
Note that as we made no special efforts to ensure that the computational workload
of the cluster was completely  identical during each run, these numbers should be treated with a certain amount
of caution -- they are reasonably indicative of the computational performance of the different 
approaches, but could easily be uncertain at the 10--20\% level.
Nevertheless, some clear trends are apparent. The three approaches that attempt to model
the carbon chemistry with only one or two additional rate equations (NL97, KC08e and KC08n)
are clearly the fastest, but do not differ significantly amongst themselves in terms of required
runtime. This likely just reflects the fact that the time taken up by the chemistry in these simulations
does not dominate the total computational cost.

\begin{table}
\caption{Computational performance of the various approaches \label{tab:perf}}
\begin{tabular}{lcc}
\hline
Method & \multicolumn{2}{c}{Approximate runtime (CPU hours)} \\
& Run 1 & Run 2 \\
\hline
G10g & 1190 & 1360 \\
G10ng & 880 & 1240 \\
NL97 & 110 & 140  \\
NL99 & 310 & 460  \\
KC08e &  100 & 130 \\
KC08n & 120  & 150 \\
\hline
\end{tabular}
\end{table}

The NL99 model is approximately a factor of three slower than these simpler models, 
reflecting its significantly greater complexity, but is a factor of three faster than the even
more complex G10g and G10ng models. The slowdown as we go from NL99 to G10g or 
G10ng is roughly in line with what we expect, given an $N_{\rm spec}^{3}$ scaling for
the cost of the chemistry: $N_{\rm spec} = 10$ in the NL99 model (nine non-equilibrium
species, plus the internal energy), while $N_{\rm spec} = 15$ in G10g and G10ng 
(14 non-equilibrium species, plus the internal energy), and so the expected slowdown
is a factor of $15^{3} / 10^{3} \simeq 3.4$. On the other hand, the slowdown between
e.g.\ NL97 and NL99 is much smaller than we might expect, given that $N_{\rm spec}
= 4$ in NL97, consistent with the chemistry not being the dominant cost in the simplest
models. Finally, the difference in runtimes between models G10g and G10ng is likely 
due to the fact that the grain-surface recombination rate coefficients are somewhat 
costly to calculate, at least if one uses the \citet{wd01} fitting functions, and their 
calculation has not yet been significantly optimized within the current version of the code.

In terms of the limits that the requirement of following the chemistry places on the size of
simulation that can be performed, it is worth bearing in mind that a factor of two increase
in spatial resolution in a three-dimensional ideal MHD Eulerian simulation will generally 
lead to a factor of sixteen increase in runtime: the number of resolution elements increases 
by a factor of eight, while the Courant condition causes the maximum timestep to decrease
by a factor of two, so that twice as many timesteps are required to reach the same physical
time. In simulations where the chemistry is the dominant computational cost, and is also
subcycled (i.e.\ evolved on a timestep smaller than the magnetohydrodynamical
timestep), then this last factor of two increase can often be avoided, since the change in
spatial resolution does not directly affect the number of chemical substeps that must be
taken. However, even in this case, the bottom line is that a factor of two improvement
in spatial resolution will lead to an order of magnitude increase in runtime. Therefore,
the difference in performance between the simplest and most complex chemical
networks is roughly equivalent to a factor of two in spatial resolution: a simulation
performed with $256^{3}$ zones using model G10g  will take roughly as long as a 
simulation performed with $512^{3}$ zones using one of models NL97, KC08e or
KC08n.  
 
\section{Results} 
\label{acc}
\subsection{Chemical abundances: time evolution}
\label{time_evol}
Having discussed the relative computational performance of the various approaches,
and shown that, as expected, the simpler models are considerably faster than the more
complex ones, we now look in more detail at the behaviour they predict for the 
chemical abundances. We begin with some of
the simplest quantities that we might expect the models to be able to reproduce, the total
masses of atomic carbon and CO formed in the simulation as a function of time. A 
convenient way to express this is in terms of the mass-weighted mean abundances of 
these species.  We can define the mass-weighted mean abundance of  a species p as
\begin{equation}
\mwfrac{\rm p} = \frac{\sum_{i,j,k} x_{\rm p}(i,j,k) \rho(i,j,k) \Delta V(i,j,k)}{M_{\rm tot}},
\end{equation}
where  $x_{\rm p}(i,j,k)$ is the fractional abundance (by number, relative to the number
of hydrogen nuclei) of species m, $\rho(i,j,k)$ is the mass density in zone $(i,j,k)$, 
$\Delta V(i,j,k)$ is the volume of zone $(i,j,k)$, $M_{\rm tot}$ is the total mass of 
gas present in the simulation, and where we sum over all grid zones. 
It is simple to convert from $\mwfrac{\rm p}$ to $M_{\rm p}$ (the total mass of species p in 
the simulation) using the following equation:
\begin{equation}
M_{\rm p} = \frac{m_{\rm p}}{x_{\rm H} m_{\rm H} + x_{\rm He} m_{\rm He}} M_{\rm tot} \mwfrac{\rm p},
\end{equation}
where $m_{\rm H}$ is the mass of a hydrogen atom and $m_{\rm He}$ is the mass of a helium
atom.

\begin{figure}
\includegraphics[height=3.4in]{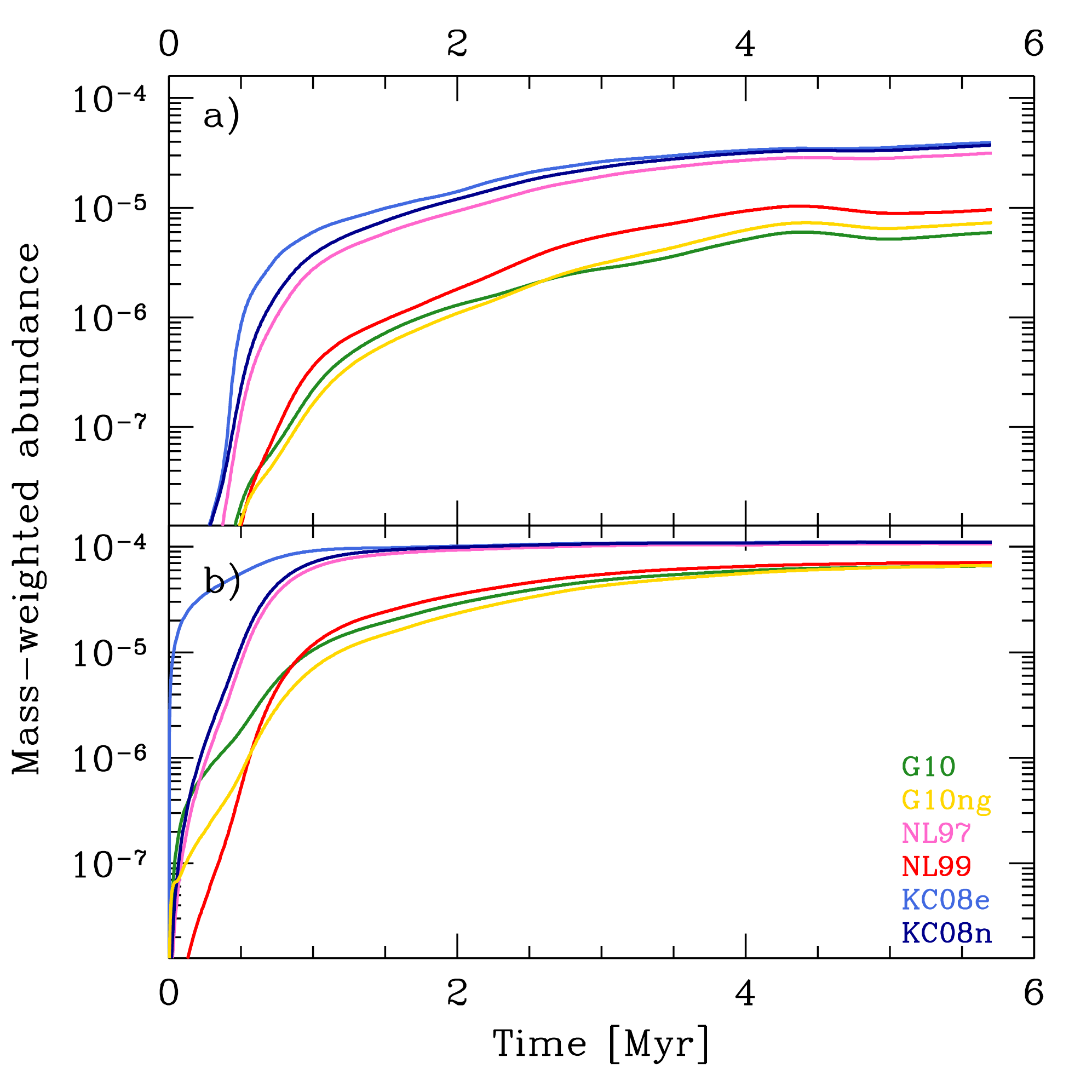}
\caption{(a) Time evolution of the mass-weighted mean CO abundance in
the simulations in set 1, representing the evolution of a low density
cloud with mean hydrogen number density $n_{0} = 100 \: {\rm cm^{-3}}$.
(b) As (a), but for the simulations in set 2, which model the evolution of
a higher density cloud, with $n_{0} = 300 \: {\rm cm^{-3}}$. 
\label{CO-time-evol}}
\end{figure}

\begin{figure}
\includegraphics[height=3.4in]{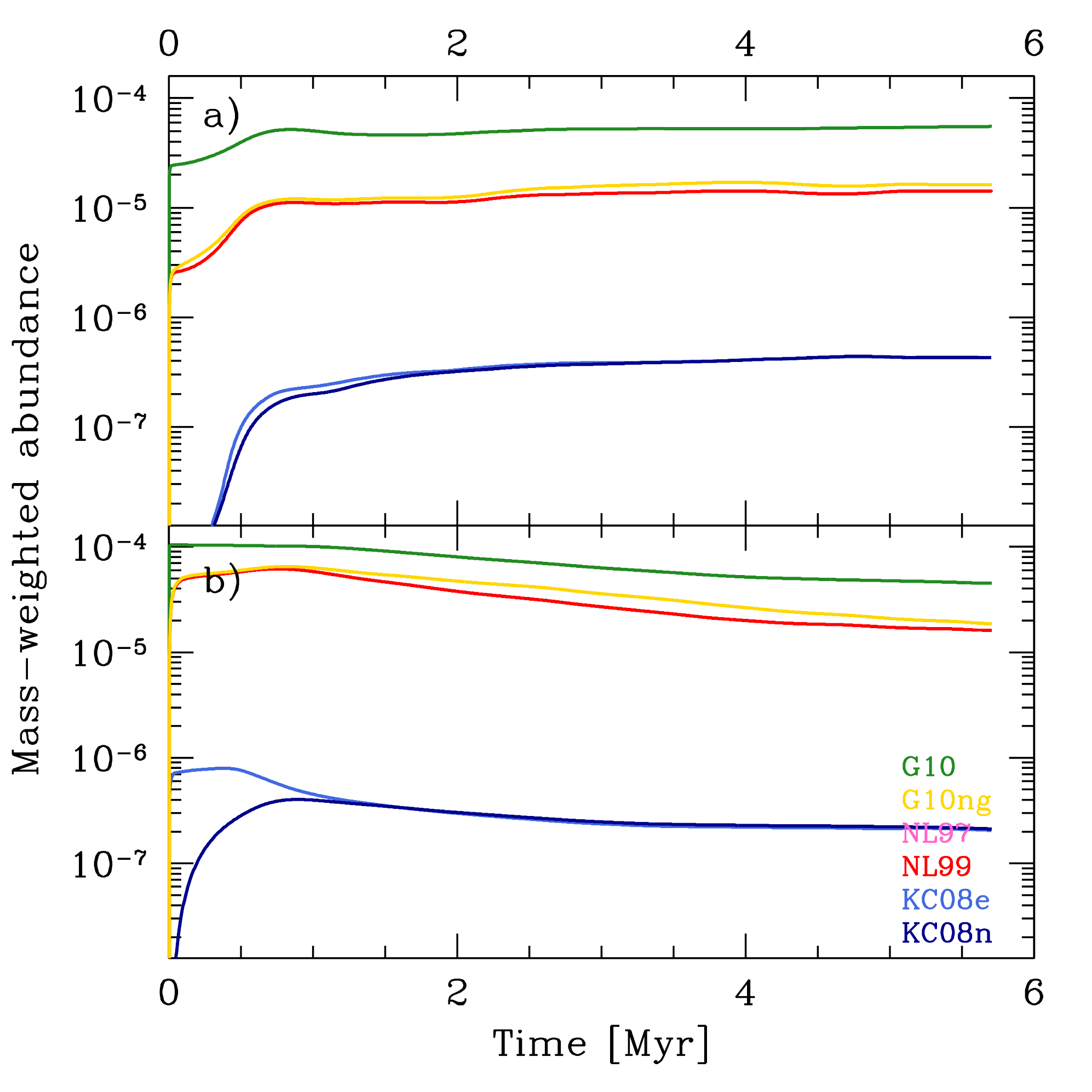}
\caption{(a) Time evolution of the mass-weighted mean abundance of
atomic carbon in the simulations in set 1, the low density cloud. Note 
that no line is plotted for the NL97 model, as this model does not track 
the abundance of atomic carbon.
(b) As (a), but for the simulations in set 2.
\label{CI-time-evol}}
\end{figure}

In Figure~\ref{CO-time-evol}a, we plot the evolution with time of the mass-weighted mean
CO abundance, $\mwfrac{CO}$, in the six simulations in set 1, our low density simulations. 
It is immediately
obvious from this figure that the different methods do not agree on the amount of CO that
is formed in the gas. Although there is good agreement between closely related models 
(e.g.\ KC08n and KC08e, or G10g and G10ng), there is roughly an order of magnitude
difference in the CO fractions predicted by the full set of models.
Furthermore, it is clear that the three simplest models (NL97, KC08e and KC08n) find a 
qualitatively different evolution for the mean CO abundance that the three more complex
models (NL99, G10g and G10ng), predicting a more rapid rise in $\mwfrac{CO}$ and a
significantly larger final value.

The behaviour of $\mwfrac{CO}$ in our higher density runs in set 2, illustrated 
in Figure~\ref{CO-time-evol}b, further supports our finding that we can divide our models
into two subsets that predict qualitatively different behaviour for the growth of the CO mass
fraction. Runs NL97, KC08e and KC08n form almost all of their CO at $t < 1 \: {\rm Myr}$,
and thereafter predict that $\mwfrac{CO}$ should barely evolve. On the other hand, models
NL99, G10g and G10ng find that CO forms over a more extended period, and predict
that $\mwfrac{CO}$ reaches a statistical steady state only at $t > 5 \: {\rm Myr}$. Runs
NL97, KC08e and KC08n agree very well on the final amount of CO produced, predicting
values for $\mwfrac{CO}$ that differ by only a few percent.
Runs NL99, G10g and G10ng also agree well at times $t > 1 \: {\rm Myr}$ on the amount of CO 
produced, but predict a final value for $\mwfrac{CO}$ that is only 60\% of the size of that 
predicted by runs NL97, KC08e and KC08n. The only difference between runs G10g and
G10ng is the inclusion of grain-surface recombination in the former, and the good agreement
between the CO distributions produced by these two models therefore indicates that the
inclusion of this process has only a minor effect on the CO abundance. The good agreement
between the results of these two models and the NL99 model is more of a surprise, and 
suggests that the composite species CH$_{x}$ and OH$_{x}$ in the NL99 model do a good
job of reproducing the behaviour of the intermediate hydrocarbons and oxygen-bearing
species that are tracked directly in models G10g and G10ng.

In Figure~\ref{CI-time-evol}, we examine the evolution of the mass-weighted mean abundance
of atomic carbon, $\mwfrac{C}$, in our simulations. In this case, we compare only five models,
as the NL97 chemical model does not contain atomic carbon, and so cannot make any prediction
whatsoever about its abundance. It is immediately apparent from the Figure that there is a far
larger degree of disagreement for the C abundances than for the CO abundances, particularly 
for our low density simulations. In these, we see three different evolutionary tracks. Run G10g 
forms a large amount of atomic carbon very quickly, reaching a value of
 $\mwfrac{C} \sim 2 \times 10^{-5}$ within only a few thousand years, an interval so short that it is not well represented in the Figure. Following this, the atomic carbon abundance continues to increase for the rest of the run, but at a  much, much slower pace, reaching a final abundance $\mwfrac{C} \sim 5.5 \times 10^{-5}$ by the end of the simulation. In runs G10ng and NL99, there is also a rapid growth in the atomic carbon abundance at very early times, but in this case, this phase of rapid growth stops once $\mwfrac{C} \sim 2 \times 10^{-6}$. From this point until $t \sim 1 \: {\rm Myr}$, the atomic carbon abundance continues to increase significantly, albeit at a much slower pace than at the very beginning of the simulation, while for $t > 1 \: {\rm Myr}$, there is little further evolution of $\mwfrac{C}$. Finally, in
runs KC08e and KC08n, the atomic carbon abundance does not display the very rapid growth at
$t \ll 1 \: {\rm Myr}$ seen in the other runs. Although there is significant growth in the atomic carbon
abundance at $t < 1 \: {\rm Myr}$, the characteristic timescale is a significant fraction of a megayear,
rather than the few thousand years found in the other models, and the amount of atomic carbon formed
is much smaller. At $t > 1 \: {\rm Myr}$, there is a clear change in behaviour; the growth in $\mwfrac{C}$ slows significantly, and it appears to have reached a steady-state by the end of the simulation. The final abundance of atomic carbon is considerably smaller than in the other runs: it is roughly a factor of thirty smaller than in than in runs G10ng or NL99, and more than a factor of 100 smaller than in run G10g.

In our higher density simulations, the size of the disagreement between run G10g and runs G10ng and NL99 
is somewhat smaller, although significant differences remain. As in the lower density case, we see a
very rapid growth in the atomic carbon abundance at very early times in these models, followed in this case
by a slow decline as the atomic carbon is incorporated into CO molecules. The KC08e model also 
predicts a rapid rise in $\mwfrac{C}$ followed by a slow decline, but the amount of atomic carbon produced
in this model is a factor of 100 or so smaller than in the other models. Finally, model KC08n predicts a 
somewhat slower rise in $\mwfrac{C}$ at early times, with a characteristic timescale of about 1~Myr, 
followed by a decline in the atomic carbon fraction at times $t > 1 \: {\rm Myr}$ that is almost indistinguishable 
from that in the KC08e model.

These discrepancies in the behaviour of the atomic carbon abundance in the various models are
actually relatively easy to understand. In the KC08n and KC08e models, atomic carbon is produced
only as an outcome of CO photodissociation, rather than directly from C$^{+}$ by recombination.
Since the formation rate of CO is relatively slow, relying as it does on a radiative association 
reaction, this means that the equilibrium abundance of atomic carbon is very small when the
visual extinction is small. Therefore, in the KC08e model, the growth of the carbon abundance
is regulated by the appearance of regions with high visual extinctions. In run 1, regions with a large 
enough visual extinction to allow for the production of non-negligible amounts of C and CO are 
created by the turbulent restructuring of the gas on a timescale of roughly $1 \: {\rm Myr}$. In run 2,
on the other hand, the higher mean density allows some regions to have a significant visual extinction
(and hence a high equilibrium abundance of atomic carbon) even at $t = 0$. In the KC08n model,
the same consideration applies with regards to the dependence on visual extinction, but in addition, the 
growth of the atomic carbon abundance is also regulated by the time taken to form CO, which itself is
of the order of $1 \: {\rm Myr}$ or longer in much of the gas. 

In the other three models, the most effective way to convert C$^{+}$ to C is by direct 
recombination: gas-phase recombination in models G10ng and NL99, and a mix
of gas-phase and grain-surface recombination in model G10g. The recombination time 
of fully ionized carbon in gas with $n = 100 \: {\rm cm^{-3}}$ and $T = 60 \: {\rm K}$ is
roughly 0.2~Myr, assuming that the ionized carbon itself is the dominant source of the 
required electrons, and since the equilibrium C/C$^{+}$ ratio at $t = 0$ is much smaller
than unity, it requires only a small fraction of a recombination time for the C/C$^{+}$ ratio 
to reach its equilibrium value. Subsequent
changes in the atomic carbon abundance are driven by two main effects: the restructuring
of the gas by the turbulence, which puts more of the gas mass into dense, high $A_{\rm V}$
regions where the equilibrium C/C$^{+}$ ratio is larger, and the conversion of C into CO
in this same dense, well-shielded gas. This latter process is responsible for the fall-off in the
atomic carbon abundance in the high density versions of runs G10g, G10ng and NL99 that
occurs at $t > 1 \: {\rm Myr}$ (see Figure~\ref{CI-time-evol}). The larger C/C$^{+}$ ratio
that we find in run G10g compared to runs G10ng and NL99 is an obvious consequence 
of the inclusion of grain-surface recombination, which is particularly effective when the
ratio of the UV field strength $G_{0}$ to the mean density $n_{0}$ is small, as it is in our
simulations. 

Finally,  we note that although it would be simple to investigate the evolution of the total mass of C$^{+}$ 
in a similar fashion to our analysis of C and CO above, we have chosen to omit this, as in 
practice, C$^{+}$, C and CO between them contain almost all of the available carbon in 
models G10g, G10ng and NL99 (and must contain 100\% of it in the other models), and so 
there is nothing new to be learned from studying the time evolution of the C$^{+}$.

\subsection{Chemical abundances: dependence on density and visual extinction}
We can gain more insight into the differences between the various models tested here 
by looking in more detail at their predictions for the distribution of CO at the end of the 
simulations. We know from previous work \citep{glo10} that regardless of whether we
examine them as a function of density or visual extinction, we always find considerable 
scatter in the CO abundances in a turbulent cloud. The reason for this is that the CO abundance
is sensitive to both density and visual extinction, but these two quantities are only poorly
correlated within turbulent clouds. Nevertheless, despite this scatter it can still be useful
to examine averaged quantities, such as the total fraction of carbon represented by 
C$^{+}$, C or CO, and how this varies as a function of density or visual extinction. 

\begin{figure}
\includegraphics[height=3.4in]{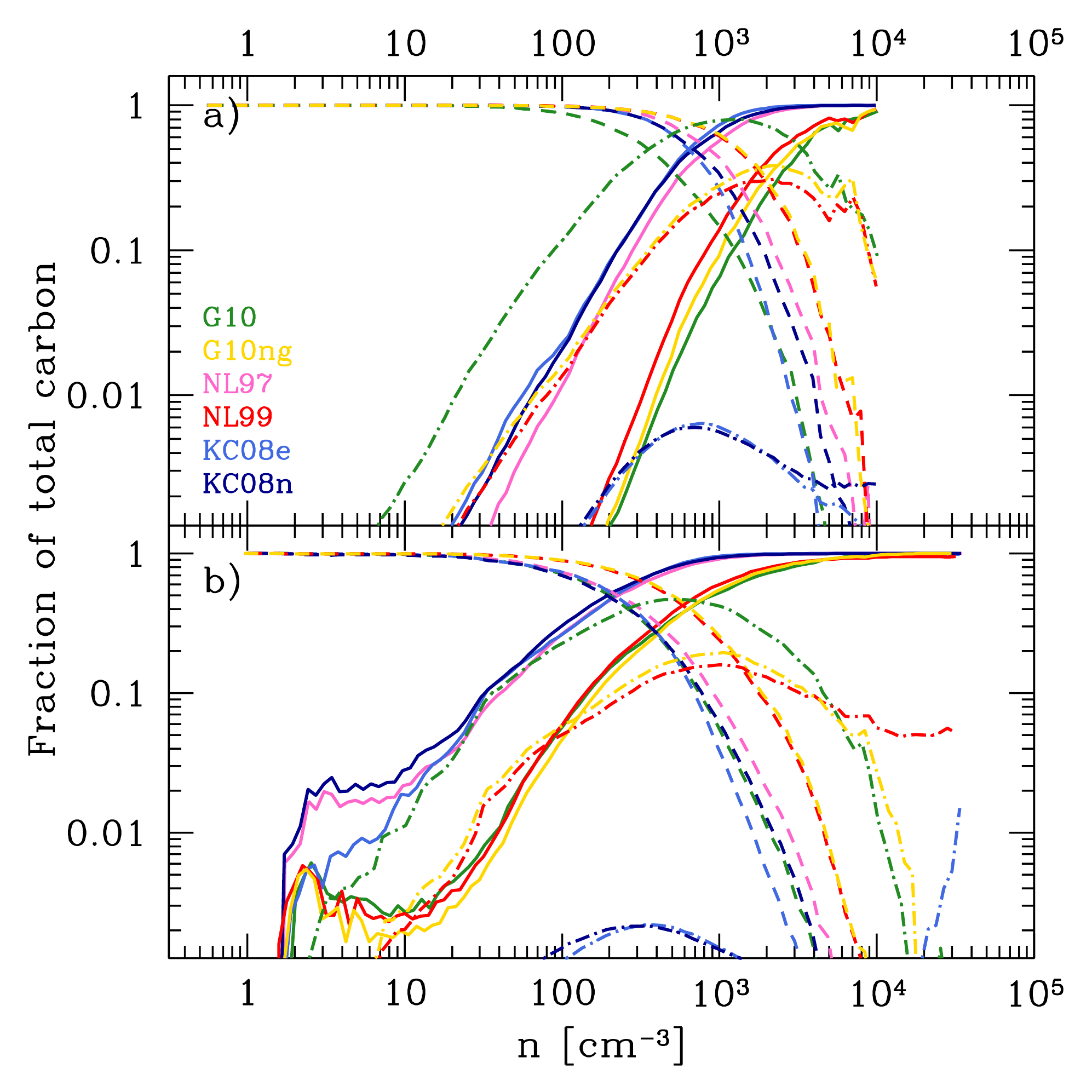}
\caption{(a) Fraction of the total available carbon found in the form of C$^{+}$ (dashed lines), 
C (dash-dotted lines), or CO (solid lines) in the simulations in set 1, plotted as a function of
density. Note that as model NL97 does not include atomic carbon, no line is plotted in this
case. (b) As (a), but for the simulations in set 2.
\label{cmf-n}}
\end{figure}

Therefore, in Figure~\ref{cmf-n}, we show how the fraction of carbon in the form of 
C$^{+}$, C or CO varies with density in the simulations. Let us focus initially on 
Figure~\ref{cmf-n}a, which shows the results from the simulations in set 1. At 
densities below the mean density of $100 \: {\rm cm^{-3}}$, all of the models predict
that almost all of the carbon will be C$^{+}$, and hence agree very well in their
predictions of the C$^{+}$ abundance. However, when we look at the abundances of
C and CO at these low densities, we find substantial disagreement between the
results of the different models. Let us start by considering the abundance of atomic
carbon. This is entirely absent in the NL97 model, but represents 0.1\% of the total carbon at
$n = 100 \: {\rm cm^{-3}}$ in the KC08 models, roughly 1\% in models G10ng and NL99,
and closer to 10\% in model G10g. The differences in the predictions for the atomic
carbon abundance coming from various models persist over a wide range of densities,
and at high densities, where atomic carbon is abundant, they also cause significant 
differences in the C$^{+}$ fraction, which, for instance, falls off more rapidly with 
increasing density in run G10g than in the other runs. Turning to the CO fraction, we
find that runs G10g, G10ng and NL99 agree well with each other on the behaviour of
the CO fraction, as do runs NL97, KC08e and KC08n, but that there is a significant
difference in the behaviour of these two sets of runs. All of the models predict a sharp
rise in the CO fraction with increasing number density, and agree that the fraction of
carbon in CO should be close to 100\% for number densities of order $10^{4} \:
{\rm cm^{-3}}$. However, at densities below $10^{4} \: {\rm cm^{-3}}$, we find that
the CO fraction in runs NL97, KC08e and KC08n is systematically larger than in the
other three runs. For example, at $n = 1000 \: {\rm cm^{-3}}$, the three simple models
predict a CO fraction of roughly 80--90\%, while runs G10g, G10ng, and NL99 predict a
value that is closer to 10\%. 

In Figure~\ref{cmf-n}b, which shows the behaviour of the simulations in set 2, we 
find very similar results. In this case, the transition from C$^{+}$ to C to CO takes place
over a wider range of densities, and the gas becomes CO dominated at a lower density
than before. This is a consequence of the higher mean extinction of the gas  -- in these
simulations, there is more gas at all densities that is well shielded from the UV 
background and hence can maintain a high CO abundance. Nevertheless, the qualitative
behaviour of the CO fraction as a function of density remains the same as in our lower density
runs. Models NL97, KC08e and KC08n continue to agree well over the whole range of densities 
plotted, as do models G10g, G10ng and NL99, with the first set of models predicting systematically
higher CO fractions than the second set. At $n > 5000 \: {\rm cm^{-3}}$, all six models agree that
the gas should be completely CO-dominated.

\begin{figure}
\includegraphics[height=3.4in]{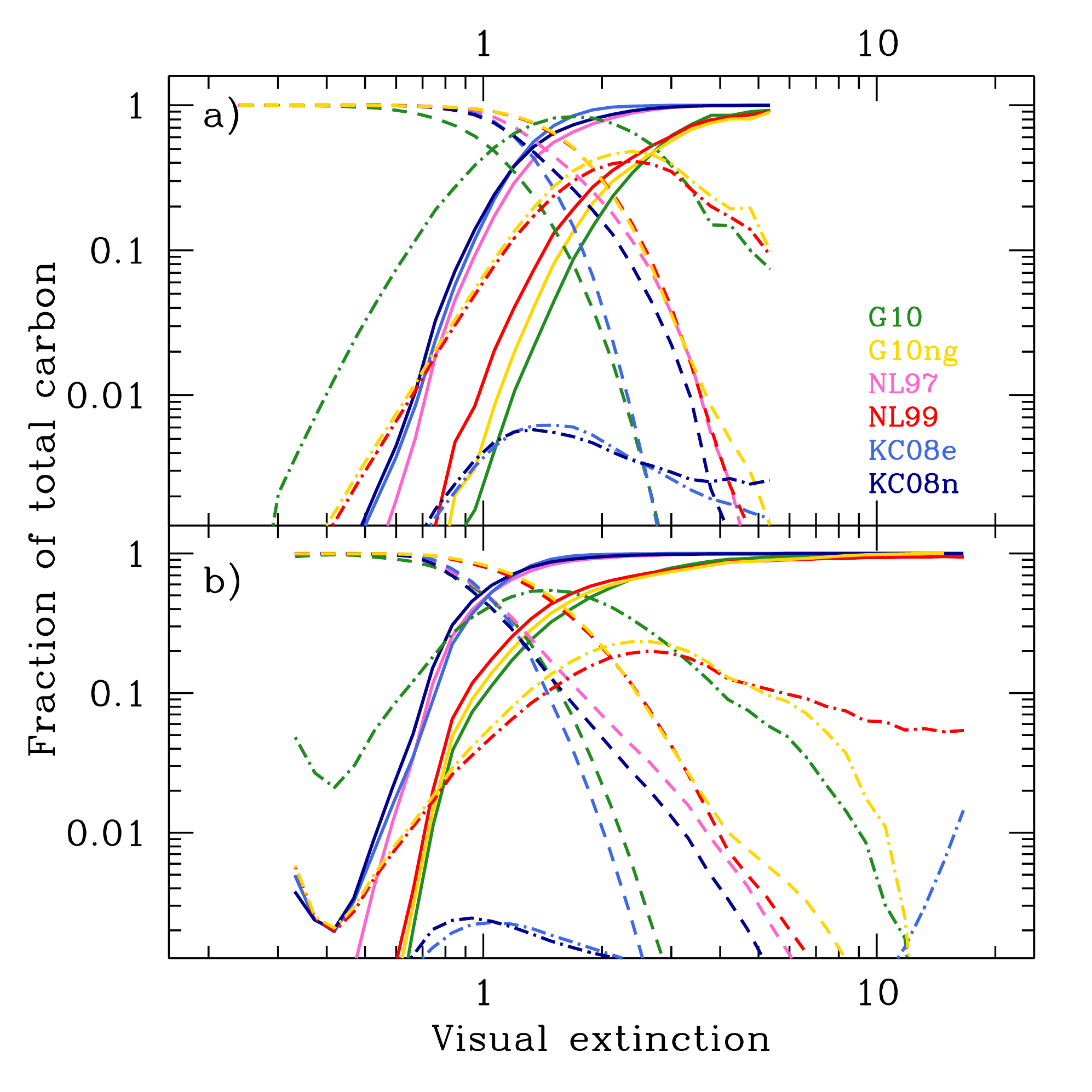}
\caption{(a) As Figure~\ref{cmf-n}a, but showing the variation in the C$^{+}$, C and CO
fractions as a function of visual extinction $A_{\rm V}$ for the simulations in set 1.
(b) As (a), but for the simulations in set 2. \label{cmf-AV}}
\end{figure}

In Figure~\ref{cmf-AV}, we show how the fraction of carbon in the form of  C$^{+}$, C or CO 
varies as a function of the effective visual extinction, defined for any given cell in our
simulations as
\begin{equation}
A_{\rm V, eff} = - \frac{1}{2.5} \ln \left[ \frac{1}{6}
\left(\sum_{p=1}^3 e^{-2.5A_{\rm V}(x_{p+})} + e^{-2.5A_{\rm V}(x_{p-})} \right) \right]
\label{aveff}
\end{equation}
where $A_{\rm V}(x_{p+})$ is the visual extinction of material between that
cell and the edge of the volume in the positive direction along the
$x_p$ axis, and so forth. The choice of the factor of 2.5 occurs because in our models,
the CO photodissociation rate scales with the visual extinction $A_{\rm V}$ 
as $\exp(-2.5 A_{\rm V})$. The value of  $A_{\rm V, eff}$ defined in this fashion corresponds
to the visual extinction used in our code, in the context of our six-ray approximation, for 
computing  the CO photodissociation rate.

Figure~\ref{cmf-AV} displays a number of familiar features. Models G10g, G10ng and NL99
all agree well on the evolution of the CO fraction with increasing visual extinction, just as 
they did on its evolution with increasing density. Models G10ng and NL99 also agree on
the behaviour of the C and C$^{+}$ fractions at all but the highest visual extinctions, 
while model G10g predicts a much larger atomic carbon fraction at $A_{\rm V, eff} \sim 1$--2 
than in the other two models, and hence a correspondingly smaller C$^{+}$ fraction. 
Models NL97, KC08e and KC08n agree well with each other regarding the CO fraction,
but produce more CO at low visual extinctions than the other three runs.

Figure~\ref{cmf-AV}b also displays a curious feature in the plot of the atomic carbon fraction 
at very low $A_{\rm V, eff}$ in runs G10g, G10ng and NL99. Below an effective visual extinction
of roughly 0.4, the atomic carbon fraction increases with decreasing $A_{\rm V, eff}$ in these
three runs. However, further investigation shows that this feature is a numerical artifact related
to our use of the six-ray approximation. As previously discussed in \citet{glo10}, the effective
visual extinction of material right at the edge of the computational volume is higher than it
should be, owing to the poor angular sampling of the radiation field, and in practice only 
few zones very close to the edge have effective visual extinctions $A_{\rm V, eff} < 0.5$.
Because of this, when we compute the mass fractions for $A_{\rm V, eff} < 0.5$, we
are averaging over only a small number of zones, and hence are far more sensitive to the
effects of outliers than we would be at higher $A_{\rm V, eff}$. In this particular case, the
anomalous behaviour of the atomic carbon fraction is due to the contribution from a small
clump of dense gas located at the edge of the simulation volume. The higher density
of this gas enables C$^{+}$ recombination to be more effective, and allows it to have a 
higher atomic carbon abundance than the rest of the gas at this visual extinction. The effect
is more pronounced in run G10g because of the inclusion of grain-surface recombination in
that model. 

\subsection{CO column densities and integrated intensities}
\label{wco_section}
Although informative, the quantities that we have examined so far have the significant 
disadvantage that they are not observable in real molecular clouds. It is therefore 
useful to examine whether the differences between the models that we have already
discussed will lead to clear differences in the observable properties of the CO
distribution.

We begin by examining the behaviour of the CO column density distribution. CO column
densities are directly observable in the ISM along low extinction ($A_{\rm V} \sim 2$ or
less) sightlines to bright background stars, where they can be accurately 
determined by UV absorption  measurements \citep[see e.g.][]{shef07}.  For higher extinctions, 
this technique is no longer effective, and one typically relies on CO emission, which is an 
accurate tracer of the  CO column density only for transitions which are optically thin and 
uniformly excited \citep[see e.g.][]{pineda08,shetty10}. Since much of the $^{12}$CO emission 
coming from Galactic GMCs is optically
thick, information on the CO column density distributions in these clouds generally comes from
observations of rarer CO isotopologues such as $^{13}$CO or  C$^{18}$O, which are
optically thin over a much wider range of cloud column densities.

\begin{figure}
\includegraphics[height=3.4in]{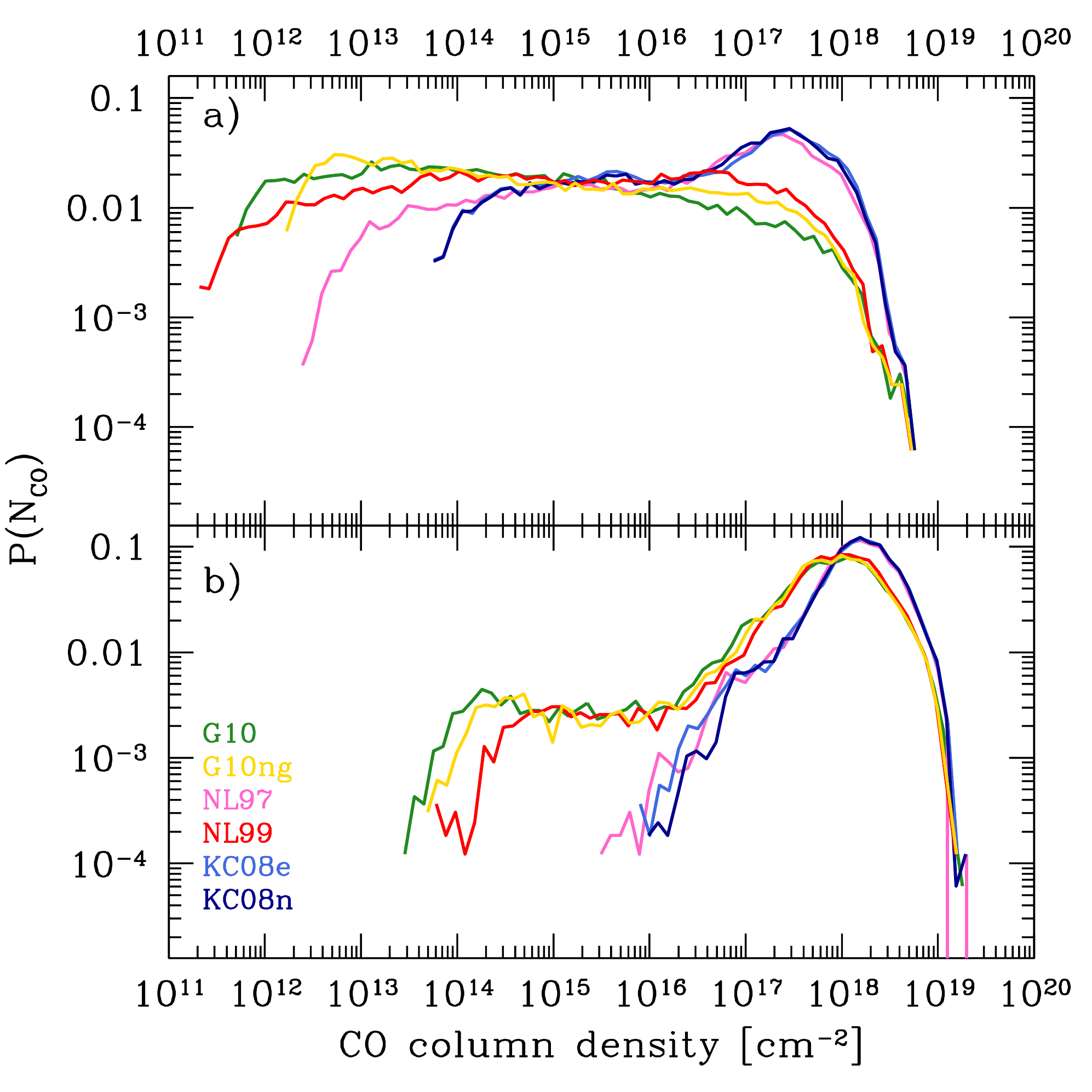}
\caption{(a) CO column density PDFs for the runs in set 1. 
(b) As (a), but for the runs in set 2.
\label{col-pdf}}
\end{figure}

The probability density functions (PDFs) for the CO column density in our two sets of runs
are plotted in Figure~\ref{col-pdf}. In our low density runs, we see that the PDFs are 
relatively flat, indicative of there being a roughly equal probability of selecting any 
particular CO column density within a relatively wide range. This is quite unlike the
PDF of total column density, which has the characteristic log-normal shape ubiquitously found 
in simulations of supersonic turbulence
\citep[see e.g.][]{osg01}, and demonstrates that the CO is not a particularly accurate tracer 
of the underlying density distribution in our low density simulations \citep[see also][for more on
this point]{shetty10}. In runs NL97, KC08e and
KC08n, we begin to see the influence of this underlying structure between CO column densities
of $10^{17} \: {\rm cm^{-2}}$ and a few times $10^{18} \: {\rm cm^{-2}}$, where there is a clear
peak in the PDF, but no such feature is visible in the CO column density distributions produced
in the other three runs. In our higher density runs, on the other hand, the influence of the
underlying density structure of the gas is far more pronounced, with all of the runs showing a
clear peak in the PDF at high CO column densities. The PDFs can best be understood as
the superposition of two different features: a log-normal portion at high $N_{\rm CO}$, corresponding 
to lines of sight along which most of the carbon is in molecular form, leading to a CO column density 
that simply traces the total column density, and an extended tail at much lower $N_{\rm CO}$ that 
corresponds to lines of sight along which the CO is significantly photodissociated.

Comparing the predictions of the different models, we see that once again the runs can be
separated into two distinct sets. Runs G10g, G10ng and NL99 agree reasonably well with each other, 
particularly for the higher density cloud, but produce results that differ significantly from those found
in runs NL97, KC08e and KC08n. The latter models produce narrower CO column density 
PDFs, with a more pronounced peak at high $N_{\rm CO}$, and with much smaller probabilities
at low $N_{\rm CO}$.  One consequence of this is that the CO is a better tracer of the underlying
density distribution in  runs NL97, KC08e and KC08n than in runs G10g, G10ng and NL99.

We have also computed estimates of the frequency-integrated intensities of the
$J = 1 \rightarrow 0$ line of CO corresponding to these CO column densities, 
using the same technique as in \citet{gm10}. To briefly summarize, we assume
that the CO is in local thermodynamic equilibrium (LTE), and is isothermal, with a 
temperature equal to a weighted mean temperature for the gas, computed using 
the CO number density as the weighting function:
\begin{equation}
T_{\rm mean, CO} = \frac{\sum_{i,j,k} T(i,j,k) n_{\rm CO}(i,j,k)}{\sum_{i,j,k} n_{\rm CO}(i,j,k)},
\end{equation}
where we sum over all grid zones $i,j,k$. We also assume that the CO linewidth is
uniform, and is given by $\Delta v = 3 \: {\rm km} \: {\rm s^{-1}}$ (which is roughly
equal to the one-dimensional velocity dispersion we would expect to find in a gas 
with an RMS turbulent velocity of $5 \: {\rm km \: s^{-1}}$). Given these assumptions,
we can relate the CO column density to the optical depth in the $J = 1 \rightarrow 0$ line
using \citep{tielens05}
\begin{equation}
\tau_{10} = \frac{A_{10} c^{3}}{8 \pi \nu_{10}^{3}} \frac{g_{1}}{g_{0}} f_{0} \left[
1 - \exp\left(\frac{-E_{10}}{kT} \right) \right] \frac{N_{\rm CO}}{\Delta v},
\end{equation}
where $A_{10}$ is the spontaneous radiative transition rate for the 
$J = 1 \rightarrow 0$ transition, $\nu_{10}$ is the frequency of the transition,
$E_{10} = h\nu_{10}$ is the corresponding energy, $g_{0}$ and $g_{1}$ are
the statistical weights of the $J=0$ and $J=1$ levels, respectively, and
$f_{0}$ is the fractional level population of the $J=0$ level. We can then
convert from $\tau_{10}$ to the integrated intensity, $W_{\rm CO}$, using 
the same curve of growth analysis as in \citet{pineda08}:
\begin{equation}
W_{\rm CO} = T_{\rm b} \Delta v \int_{0}^{\tau_{10}} 2 \beta(\tilde{\tau}) {\rm d}\tilde{\tau}, 
\label{wco}
\end{equation}
where $T_{\rm b}$ is the brightness temperature of the line, which we assume is
simply equal to $T_{\rm mean, CO}$, and $\beta$ is the photon escape probability,
given by 
\begin{equation}
\beta(\tau) = \left \{ \begin{array}{lr} [1 - \exp(-2.34\tau)] / 4.68 \tau & \tau \leq 7, \\
\left(4\tau \left[\ln \left(\tau / \sqrt{\pi} \right) \right]^{1/2} \right)^{-1} & \tau > 7.
\end{array} \right.
\label{beta}
\end{equation}
Although the estimates of $W_{\rm CO}$ generated by this procedure are probably
accurate to within only a factor of a few, this level of accuracy is sufficient for our
purpose here.

\begin{figure}
\includegraphics[height=3.4in]{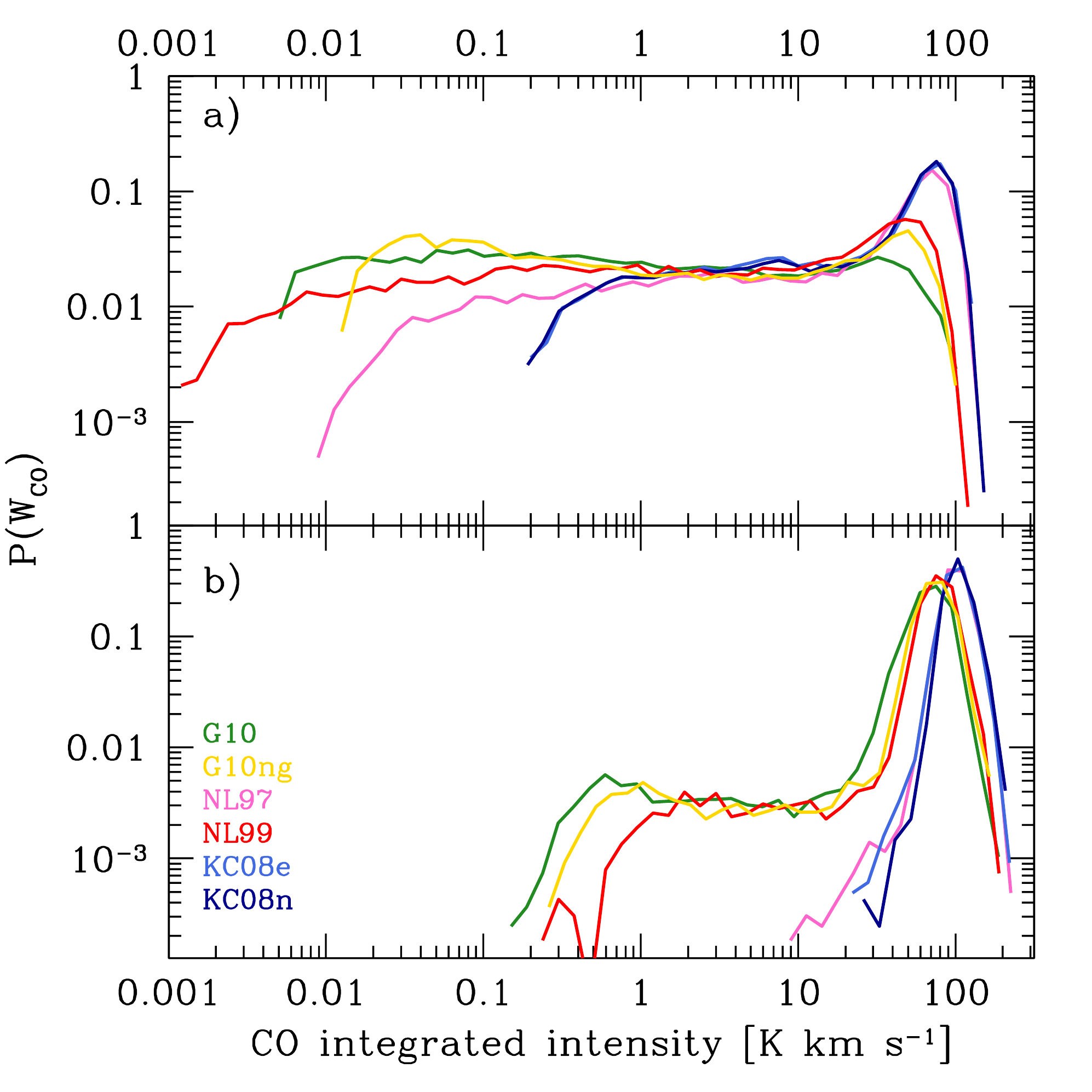}
\caption{(a) PDF of the integrated intensity in the $J = 1 \rightarrow 0$ line of
$^{12}$CO in the runs in set 1. The integrated intensities are estimates, calculated
using the technique described in \citet{gm10} and summarized in Section~\ref{wco_section}.
(b) As (a), but for the runs in set 2. \label{wco-pdf}} 
\end{figure}

In Figure~\ref{wco-pdf}, we plot PDFs of integrated intensity for the simulations. At $W_{\rm CO}
< 30 {\rm K \: km \: s^{-1}}$, these PDFs closely resemble the CO column density PDFs, while at
higher integrated intensities, we see a clear peak in the PDF even when there is no corresponding
feature in the CO column density distribution at high $N_{\rm CO}$. This behaviour 
can be easily understood as being due to the change from having optically thin CO emission at 
$W_{\rm CO} \sim 30 {\rm K \: km \: s^{-1}}$ and below, to having optically thick emission at higher
$W_{\rm CO}$. In the optically thin regime, $W_{\rm CO} \propto N_{\rm CO}$, and so we see 
similar behaviour in both PDFs. In the optically thick regime, on the other hand, $W_{\rm CO}$
increases only slowly with increasing $N_{\rm CO}$, and so we find a `pile-up' of values at 
these intensities \citep[see also the more detailed discussion of this effect in][]{shetty10}.

\subsection{The C/CO ratio}
Another potentially useful quantity for distinguishing between different models is the 
ratio of the column density of atomic carbon to that of CO. A number of groups have attempted
to measure this value \citep[see e.g.][]{ingalls97,ikeda02,ben03}, typically finding values in
the range of 0.1 to 3, with the higher values coming from lower column density translucent
clouds, and the lower values coming from higher column density dark clouds. Although a
comprehensive comparison between our results and these observational determinations is
outside of the scope of our current paper, a simple comparison nevertheless proves 
illuminating. 

In Table~\ref{tab:CCO}, we list the ratio of the mean column density of atomic
carbon, $\bar{N}_{\rm C}$, to the mean column density of CO, $\bar{N}_{\rm CO}$, that we
obtain for each of our runs, with the exception of the two NL97 models, which do not track
atomic carbon. We see that two of the models -- G10ng and NL99 -- produce values for the
C/CO column density ratio that are broadly in line with the values observed in real clouds.
On the other hand, model G10g in run 1 produces a significantly higher value than is observed,
suggesting that our treatment of C$^{+}$ recombination on grains may actually overestimate
the rate at which this process occurs in the real ISM. In this context, it is interesting to note
a recent study by \citet{liszt11} that also finds indications that we do not currently understand
the role that grain-surface recombination of C$^{+}$ plays in the transition from C$^{+}$ to
C to CO. Finally, we see from Table~\ref{tab:CCO} that the KC08e and KC08n models 
produce significantly smaller C/CO ratios than are seen in real clouds, likely because these
models neglect the effect of gas-phase recombination of C$^{+}$.

\begin{table}
\caption{Comparison of C/CO column density ratios \label{tab:CCO}}
\begin{tabular}{lcc}
\hline
Method & \multicolumn{2}{c}{$\bar{N}_{\rm C} / \bar{N}_{\rm CO}$} \\
\hline
& Run 1 & Run 2 \\
\hline
G10g & 9.3 & 0.68 \\
G10ng & 2.2 & 0.28 \\
NL99 & 1.5 & 0.23 \\
KC08e &0.01 & 0.002 \\
KC08n & 0.01 & 0.002 \\
\hline
\end{tabular}
\end{table}

\subsection{The CO-to-H$_{2}$ conversion factor}
Before we conclude our study of the details of the CO distribution in our simulations, it is 
interesting to examine whether the differences in the amount of CO formed in the various
runs, and in its spatial distribution, lead to significant differences in the CO-to-H$_{2}$
conversion factor, $X_{\rm CO}$. This is conventionally defined as the ratio of the H$_{2}$
column density to the integrated intensity of the $J = 1 \rightarrow 0$ transition of CO, i.e.\
\begin{equation}
X_{\rm CO} = \frac{N_{\rm H_{2}}}{W_{\rm CO}}.
\end{equation}
Observations of Galactic GMCs show that $X_{\rm CO}$ appears to be roughly constant 
from cloud to cloud, with a mean value of  $X_{\rm CO} =  2 \times 10^{20} {\rm cm^{-2} \:
 K^{-1} \: km^{-1} \: s}$ \citep[see e.g.][]{dame01}, and in \citet{gm10} we showed that our
 turbulent cloud models reproduce this behaviour, provided that the mean visual extinction
 of the clouds exceeds a threshold value of $A_{\rm V} \sim 2$--3. 
 
 To estimate $X_{\rm CO}$ for the simulations considered in this paper, we start by finding
 the mean of the distribution of integrated intensities that we have already computed for each 
 simulation. Armed with a mean integrated intensity for each simulation, we next compute the
 mean H$_{2}$ column density for each simulation, following which we can obtain our estimate
 of  $X_{\rm CO}$ simply by taking the ratio of these two mean quantities. The values obtained
 using this procedure are listed in Table~\ref{tab:Xfactor}.
 
In the low density case, models NL97, KC08e and KC08n yield values for $X_{\rm CO}$ that are
a factor of a few smaller than the standard Galactic value, while the other three runs yield values
that are in much better agreement with the observations. However, given that our estimates of
$X_{\rm CO}$ here are likely only accurate to within a factor of a few, it is unclear whether the
difference between predictions of the NL97, KC08e and KC08n models and the observations is 
really meaningful. In any case, in the high density case we find 
much better agreement between the models, with all of them now producing values
that are somewhat smaller than the Galactic value. The reason for this similarity is that all of the runs 
agree reasonably well (to within a factor of two) on the peak value of the integrated CO intensity found 
in the simulation, with the significant differences in the $W_{\rm CO}$ distribution occurring for values
considerably below the peak. Our estimate of $X_{\rm CO}$ is dominated by the contribution of
lines of sight with integrated intensities close to the peak value, and is insensitive to the behaviour
of lines of sight with low $W_{\rm CO}$, and so this diagnostic tells us little about the low
intensity sightlines. 

\begin{table}
\caption{Comparison of the CO-to-H$_{2}$ conversion factor, $X_{\rm CO}$, at the
end of the runs \label{tab:Xfactor}}
\begin{tabular}{lcc}
\hline
Method & \multicolumn{2}{l}{$X_{\rm CO}$ [$10^{20} \: {\rm cm^{-2}} \: ({\rm K \: km \: s^{-1}})^{-1}$]} \\
\hline
& Run 1 & Run 2 \\
\hline
G10g & 2.67 & 1.30 \\
G10ng & 2.01 & 1.23 \\
NL97 & 0.57 & 0.90 \\
NL99 & 1.47 & 1.17  \\
KC08e & 0.50 & 0.89 \\
KC08n & 0.49 & 0.87 \\
\hline
\end{tabular}
\end{table}

\subsection{Gas temperature}
C$^{+}$, C and CO are all important coolants at the temperatures and densities found
within molecular clouds, and so differences in the predicted distributions of these species
may have an effect on the thermal evolution of the gas and on its temperature structure at
the end of the simulation.

\begin{figure}
\includegraphics[height=3.4in]{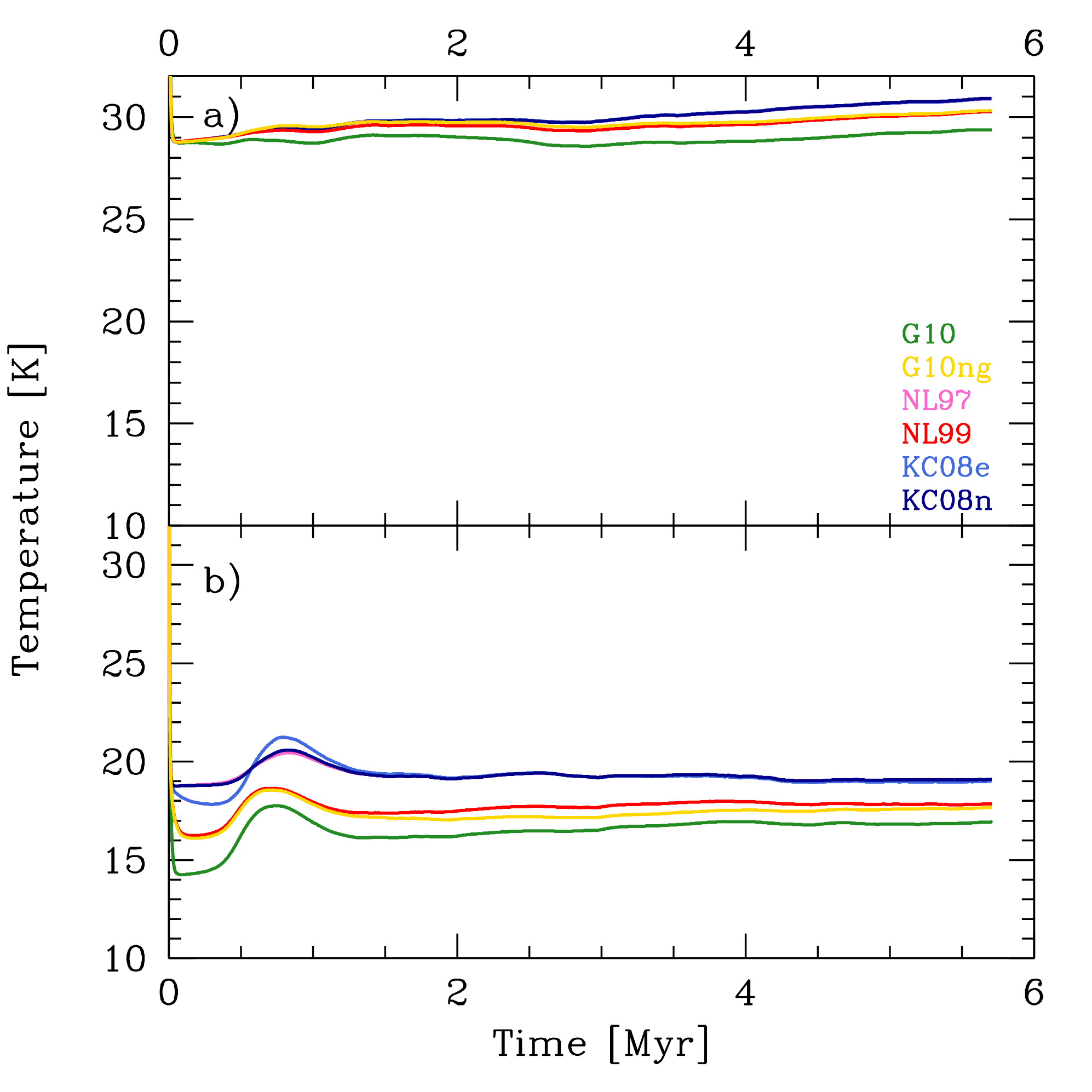}
\caption{(a) Time evolution of the mass-weighted mean gas temperature in the
simulations in set 1.
(b) As (a), but for the simulations in set 2.
\label{tmean}}
\end{figure}

We have investigated this by examining the time evolution of the mass-weighted mean
gas temperature, plotted in Figure~\ref{tmean}. In the low density runs, the mean temperature
drops very rapidly from our initial value of 60~K to a value of approximately 30~K. Thereafter,
it evolves very little over the course of the simulation. The six different chemical models
predict very similar values for the mean temperature. The most discrepant outcome is for
model G10g, which predicts a final mean temperature that is roughly 1~K cooler than in
the other runs. This result is easy to understand given the results for the mean mass-weighted
abundances of C$^{+}$, C and CO in these runs, discussed earlier in Section~\ref{time_evol}. 
C$^{+}$ is the dominant form of carbon in all of these runs, and hence will be the dominant
coolant. Differences in the abundances of C or CO will therefore have only a minor impact 
on the energy balance of the gas, and hence will have little effect on the mean gas temperature.
However, in the case of model G10g, the atomic carbon abundance becomes large enough 
that its contribution to the cooling of the gas can no longer be neglected. Since the energy
separation of the fine structure levels in neutral carbon is significantly smaller than in
singly ionized carbon, the former is a more effective low temperature coolant than the latter, and
so increasing the C abundance at the expense of the C$^{+}$ abundance leads to an overall
lower temperature for the gas. 

In the higher density runs, the mean temperature again drops very rapidly initially, before stabilizing
at a value of around 17--19~K. In this case, the difference between the models is more pronounced,
with runs NL97, KC08 and KC08n predicting the highest temperatures, and run G10g predicting the
lowest. However, at the end of the simulations, the difference between the runs is only 2~K,
or roughly 10\%. The differences can again be easily understood on the basis of our prior results
for the mean abundances. In this case, the C and CO abundances are much higher, and so  
C$^{+}$ is no longer the dominant radiative coolant. Instead, the cooling is typically dominated by 
CO (runs NL97, KC08e and KC08n) or by a mixture of atomic carbon and CO (the other three runs).
There is a clear inverse correlation between the amount of atomic carbon present in the gas and the
final mean temperature.

\begin{figure}
\includegraphics[height=3.4in]{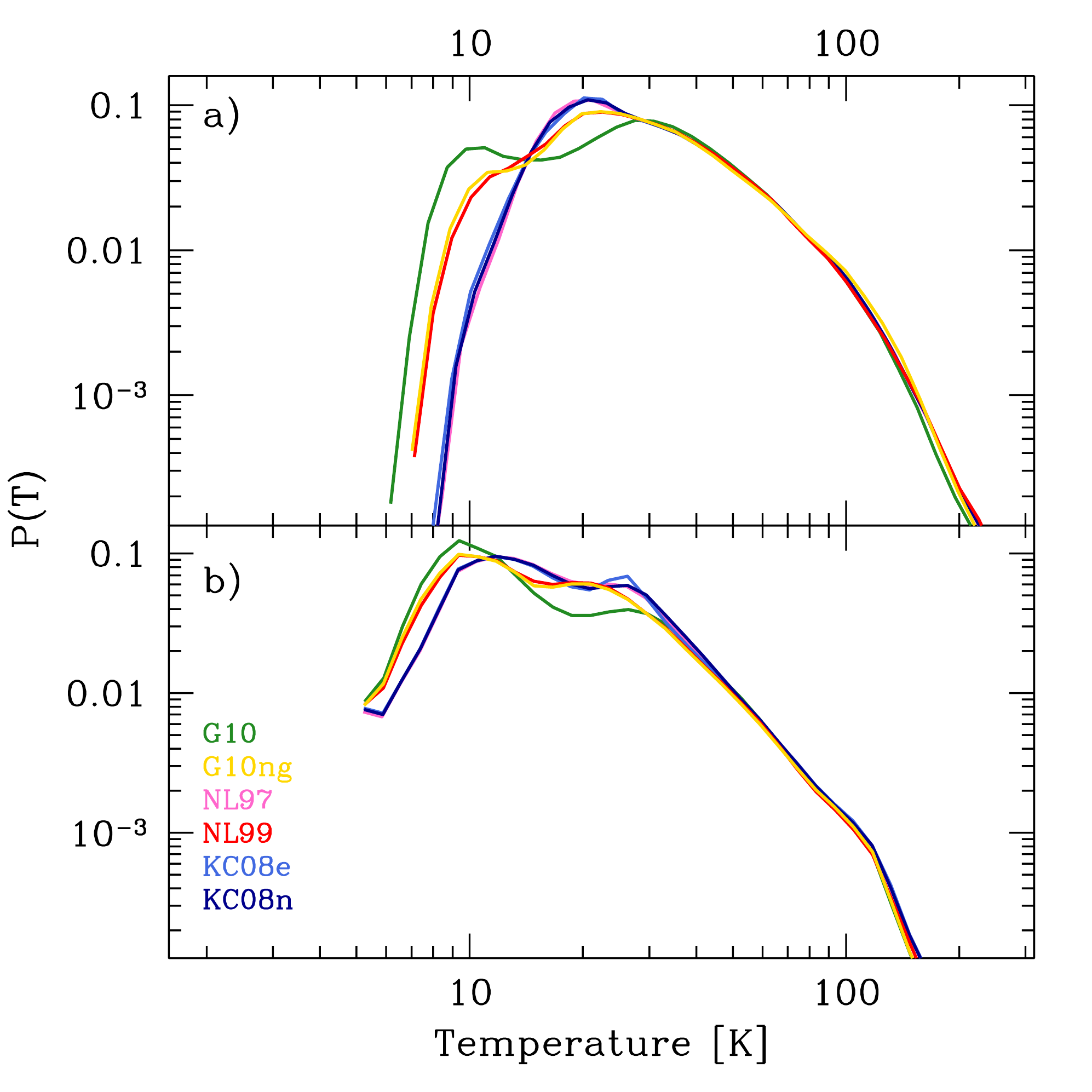}
\caption{(a) Mass-weighted temperature PDF for run 1. (b)
As (a), but for run 2. Note that in both cases, the results for the NL97
and KC08e runs are barely distinguishable from those for the KC08n
run.
\label{temp-pdf}}
\end{figure}

We have also examined the temperature distribution of the gas at the end of the simulations,
as shown in Figure~\ref{temp-pdf}, where we plot mass-weighted PDFs of the gas temperature 
for the various runs. In both the low and the high density cases, the PDF at $T > 30 \: {\rm K}$ shows almost
no dependence on the chemical model, consistent with it representing gas which is dominated
by C$^{+}$ in all of the models. Below 30~K, however, clear differences become apparent.
In the low density case, runs G10ng, NL99, KC08n, KC08e and NL97 all show a clear peak
at a temperature of around 20~K, but in run G10g, this peak occurs at the lower temperature of
10~K. Runs NL97, KC08e and KC08n all produce very little gas cooler than 10~K, but in the
other three runs we find a significant fraction of gas with temperatures in the range of 5--10~K.
In the high density run, the behaviour changes slightly. Runs NL97, KC08e and KC08n still 
have a clear feature in the PDF at $T \sim 20 \: {\rm K}$, but the true peak is now found at a
temperature slightly larger than 10~K. In runs G10g, G10ng and NL99, on the other hand, the
peak is at the slightly lower temperature of 9~K, with this feature being more pronounced in
run G10g than in the other two runs. 
The behaviour of these three runs at temperatures below this peak value is very similar.

The differences between the low temperature behaviour in the various runs can be relatively
easily understood. In the low density runs, little CO is formed, and so in most of the gas, the
dominant cooling mechanisms are C$^{+}$ and C fine structure emission, with some contribution
from dust in the densest regions. The energy separation of the $^{2}$P$_{1/2}$ ground state 
and the $^{2}$P$_{3/2}$ excited state of C$^{+}$ is approximately 92~K, and hence it cannot 
easily cool the gas below about 15--20~K, owing to the exponential suppression of the cooling rate. 
On the other hand, atomic carbon has a separation of only 24~K between its two lowest lying 
energy levels, and so remains an effective coolant down to temperatures of order 5~K. Therefore, 
chemical models producing greater quantities of atomic carbon will tend to produce lower gas 
temperatures. For instance, as we have already seen, run G10g produces more atomic carbon 
in intermediate density gas ($n \sim 1000 \: {\rm cm^{-3}}$) than runs G10ng or NL99. In the
low density case, little CO is formed, as we have already seen, and so most of the 10~K  gas
produced in this case comes from regions that are dominated by cooling from neutral carbon.
Consequently, run G10g produces significantly more of this 10~K gas than runs G10ng or NL99.
On the other hand, run NL97 produces no neutral carbon, and hence only very little 10~K gas.
In the higher density simulations, which have much higher mean CO abundances, much more
of the 10~K gas is situated in regions dominated by CO cooling, and so the differences between
runs G10g, G10ng and NL99 are much smaller than in the low density case.
The lack of neutral carbon in run NL97, and the
fact that it underproduces CO at low densities mean that it still disagrees with these three models,
and produces less very cold gas. The behaviour of runs KC08e and KC08n is very similar to that
of run NL97 because the small amount of neutral carbon produced in runs KC08e and KC08n
is not enough to significantly affect the thermal balance of the gas, and as we have already seen,
the CO and C$^{+}$ abundances in these runs are very similar to those in run NL97.

\subsection{Gas density}
Finally, we have investigated whether the differences in the temperature distributions
examined above lead to any significant differences in the density distribution of the gas,
as quantified by the density PDF. This is of particular importance for understanding the
level of accuracy in the modelling of the CO chemistry that is necessary in order to allow
us to model star formation accurately, as there are a number of theoretical models that 
argue that it is the shape of the density PDF that determines the stellar initial mass function
\citep{pnj97,pn02,hc08,hc09} or the star formation rate \citep{km05,pn09} in a molecular cloud.

\begin{figure}
\includegraphics[height=3.4in]{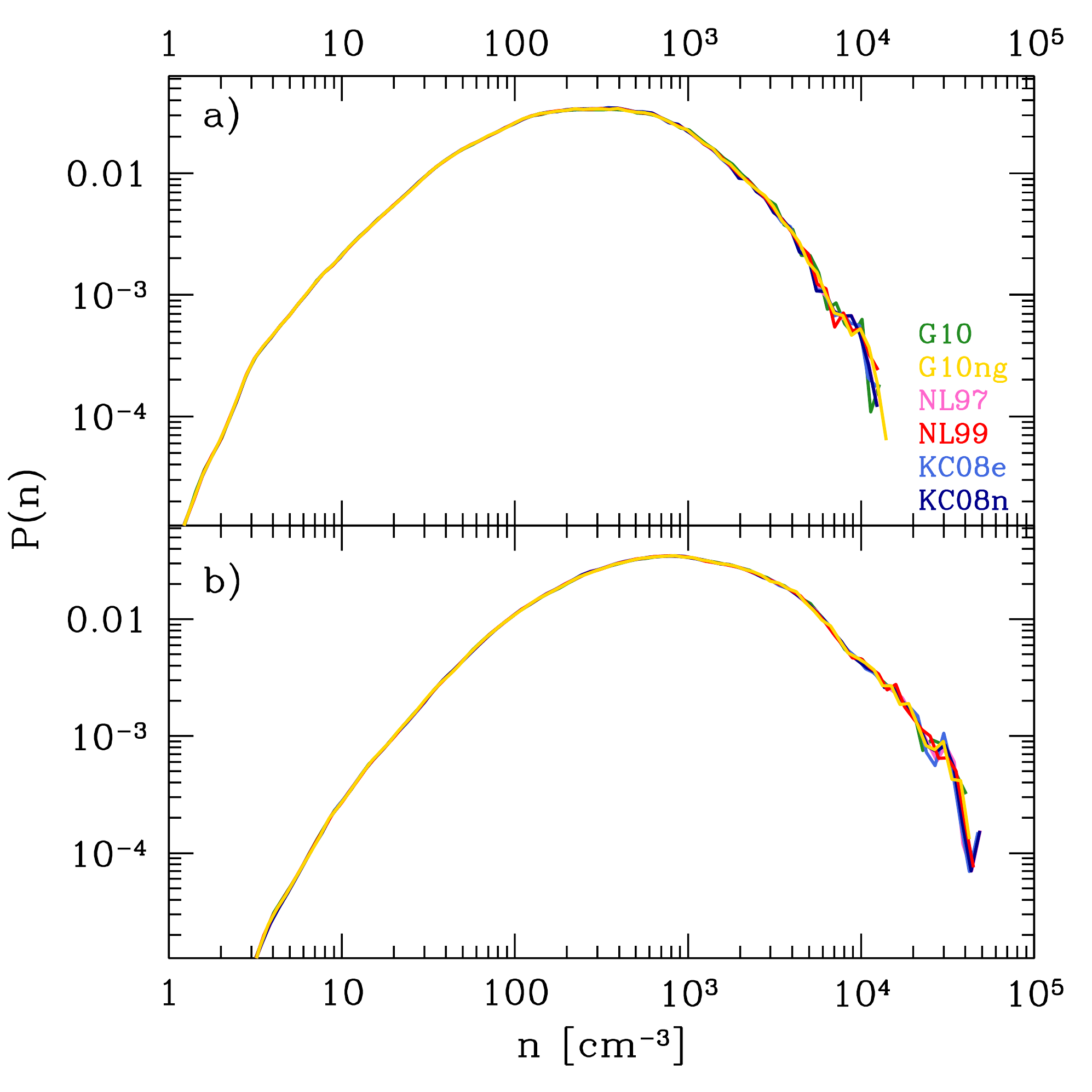}
\caption{(a) Mass-weighted density PDF for the runs in set run 1. 
Minor differences are apparent in the high density tail,  at
$n > 5000 \: {\rm cm^{-3}}$, but over the rest of the density
range, all six runs agree very well.  
(b) As (a), but for the runs in set 2. Again, there is very good
agreement between the runs at almost all densities, with only 
a few minor differences becoming apparent in the high density
tail. \label{dpdf}}
\end{figure}

In Figure~\ref{dpdf}a, we plot the density PDFs for the simulations in set 1. In Figure~\ref{dpdf}b,
we show a similar plot for the simulations in set 2. We see that in both cases, over most of the 
density range spanned by the PDF, there is essentially no difference between the results of 
any of the runs, indicating that at these densities, the physical structure of the gas is insensitive 
to the CO abundance. Minor differences start to become apparent in the high density tail of the
PDFs, but even here the differences are small. In addition, they occur at
densities that are not well resolved in our current simulations, and we cannot be completely
certain that they would persist in higher resolution simulations. 

\citet{pnj97} presented results indicative of a relationship between the dispersion $\sigma_{s}$
of the logarithmic density contrast $s = \ln (\rho / \rho_{0})$ (where $\rho_{0}$ is the mean gas
density), and the volume-weighted RMS Mach number ${\cal M}$, finding that
\begin{equation}
\sigma_{s}^{2} = \ln \left(1 + b^{2} {\cal M}^{2} \right),
\end{equation}
with $b \simeq 0.5$. Recent work by \citet{fks08} has shown that the proportionality parameter
$b$ is sensitive to the relative strength of the solenoidal and compressive modes in the forcing
field used to drive the turbulence, and ranges from $b \sim 1/3$ for purely solenoidal forcing to
$b \sim 1$ for purely compressive forcing. In each case, the effects of the gas temperature enter
through the dependence of $\sigma_{s}$ on the volume-weighted RMS Mach number. In our
simulations, the dominant contributions to ${\cal M}$ come from low density, warm gas that is 
dominated by C$^{+}$ cooling. The differences in the temperature distribution of the cold, 
denser material have little influence on ${\cal M}$. We find that ${\cal M}$ varies by no more
than 1--2\% from run to run in both the low density and high density cases, thereby explaining
why we see so little variation in the density PDF from run to run.

\section{Discussion}
\label{discuss}
\subsection{Understanding the differences in CO production}
Our comparisons in the previous section have demonstrated that the G10g, G10ng and NL99
models agree very well regarding the time evolution of the CO abundance and the spatial 
distribution of the CO. The most significant difference between the treatment of the carbon 
chemistry in these three models turns out to be the inclusion of grain-surface recombination
in model G10g, or more specifically, the recombination of C$^{+}$ ions on grain surfaces. 
The inclusion of this process has a significant effect on the ratio of C$^{+}$ to C in the gas,
and therefore indirectly affects the temperature distribution, since C is a much more
effective coolant than C$^{+}$ at gas temperatures below 20~K.

The KC08e, KC08n and NL97 models also agree well with each other, but disagree with the 
G10g, G10ng and NL99 models in two major respects. First, they predict a shorter formation
timescale for the CO, particularly in the high density run. Second, they predict systematically
larger values for the CO fraction as a function of density or visual extinction. Both of these
disagreements between the two sets of models result from the same underlying cause. 
In the KC08e, KC08n and NL97 models, the rate-limiting step in the formation of CO is the
initial radiative association between C$^{+}$ and H$_{2}$, a reaction that has a rate coefficient 
$k_{0} = 5 \times 10^{-16} \: {\rm cm^{3}} \: {\rm s^{-1}}$ in the NL97 model, and a similar value in 
the other two models. The rate per unit volume at which CH$_{2}^{+}$ ions (or CH$_{x}$ radicals) form 
in this model is therefore given by
\begin{equation}
R_{1} = 5 \times 10^{-16} n_{\rm C^{+}} n_{\rm H_{2}}.
\end{equation}
In the G10g, G10ng and NL99 models, on the other hand, there are multiple routes leading to
the formation of CO and hence the situation is somewhat more complicated. In practice, the
main contributions come from two main reactions pathways, one initiated by the radiative 
association of C$^{+}$ with H$_{2}$, as above, and a second initiated by the reaction of
atomic carbon with H$_{3}^{+}$. The latter reaction occurs rapidly, and the rate-limiting step
for this second reaction pathway
is hence not the reaction between C and H$_{3}^{+}$, but rather the formation of the H$_{3}^{+}$
ion itself. This occurs as a consequence of the cosmic ray ionization of H$_{2}$, which yields
H$_{2}^{+}$ ions that then rapidly react with H$_{2}$ to form H$_{3}^{+}$:
\begin{equation}
{\rm H_{2}^{+}} + {\rm H_{2}} \rightarrow {\rm H_{3}^{+}} + {\rm H}.
\end{equation}
In regions where at least a few percent of the carbon is ionized, the free electron abundance
is roughly equal to the abundance of ionized carbon, and the dominant destruction mechanism 
for the H$_{3}^{+}$ ions is dissociative recombination:
\begin{eqnarray}
{\rm H_{3}^{+}} + {\rm e} & \rightarrow & {\rm H_{2}} + {\rm H}, \\
 & \rightarrow & {\rm H} + {\rm H} + {\rm H}.
\end{eqnarray}
In these conditions, the net rate at which CH$_{2}^{+}$ ions are produced by reactions between
C and H$_{3}^{+}$ is therefore given by
\begin{equation}
R_{2} = 2 \zeta_{\rm H} n_{\rm H_{2}} \frac{k_{\rm CH_{2}^{+}}}{k_{\rm dr}} \frac{n_{\rm C}}{n_{\rm C^{+}}},
\end{equation}
where $\zeta_{\rm H}$ is the cosmic ray ionization rate of atomic hydrogen, $k_{\rm dr}$ is 
the rate coefficient for the dissociative recombination of H$_{3}^{+}$ ions and $k_{\rm CH_{2}^{+}}$
is the rate coefficient for the reaction
\begin{equation}
{\rm C + H_{3}^{+} \rightarrow CH_{2}^{+} + H}.
\end{equation}
In cold gas, $k_{\rm CH_{2}^{+}} / k_{\rm dr} \simeq 0.01$, and hence 
\begin{equation}
R_{2} \simeq 2 \times 10^{-19} \zeta_{\rm H, 17} \, n_{\rm H_{2}}  \frac{n_{\rm C}}{n_{\rm C^{+}}},
\end{equation}
where $\zeta_{\rm H, 17} = \zeta_{\rm H} / 10^{-17} \: {\rm s^{-1}}$. (Recall that in our simulations,
$\zeta_{\rm H, 17} = 1$). The net rate of formation of CH$_{2}^{+}$ ions in models G10g, G10ng and NL99 is approximately equal to $R_{1} + R_{2}$, since other processes make only minor contributions, while in models NL97, KC08e and KC08n, the rate of formation of CH$_{2}^{+}$ ions is simply $R_{1}$. 
Note, however, that since $R_{1}$ depends on $n_{\rm C^{+}}$, which is generally larger in models
NL97, KC08e and KC08n than in the other models, the size of the rate $R_{1}$ in the simple models
is generally larger than in the more complex models. To avoid confusion, it is useful to distinguish between these two cases by writing the rate in the simple models (NL97, KC08e, KC08n) as
$R_{\rm 1, s}$ and writing the rate in the more complex models (NL99, G10ng, G10g) as
$R_{\rm 1, c}$.

We next consider the conditions in which the rate of CH$_{2}^{+}$ formation in the complex models,
given by $R_{\rm 1, c} + R_{2}$, is bigger than the rate in the simple models, given by $R_{\rm 1, s}$.
Starting with the inequality,
\begin{equation}
R_{\rm 1, c} + R_{2} > R_{\rm 1, s},
\end{equation}
we can subtract $R_{\rm 1, c}$ from both sides, giving us
\begin{equation}
R_{2} > (R_{\rm 1, s} - R_{\rm 1, c}).
\end{equation}
If we denote the number density of C$^{+}$ in the simple models as $n_{\rm C^{+}, s}$, and
denote the same quantity in the complex models as $n_{\rm C^{+}, c}$, then we can write
this inequality as
\begin{equation}
2 \times 10^{-19} \zeta_{\rm H, 17} \, n_{\rm H_{2}}  \frac{n_{\rm C}}{n_{\rm C^{+}, c}}
> 5 \times 10^{-16} n_{\rm H_{2}} (n_{\rm C^{+}, s} - n_{\rm C^{+}, c}) \label{ineq}
\end{equation}
where we have assumed that the H$_{2}$ abundance is the same in both cases.
If the CO abundance is small compared to the abundances of C$^{+}$ and/or C, 
then in the simple models, the number density of C$^{+}$ ions will be roughly
equal to the number density of carbon atoms in all forms, i.e.\ $n_{\rm C^{+}, s} \simeq n_{\rm C, tot}$.
In the complex models, on the other hand, we have $n_{\rm C^{+}, c} + n_{\rm C} \simeq n_{\rm C, tot}$,
and hence in these conditions $n_{\rm C^{+}, s} - n_{\rm C^{+}, c} \simeq n_{\rm C}$. We can
use this fact to simplify Equation~\ref{ineq}, which reduces to the following constraint on
$n_{\rm C^{+}, c}$:
\begin{equation}
n_{\rm C^{+}, c} < 4 \times 10^{-4}  \zeta_{\rm H, 17}.
\end{equation}
In practice, this inequality is rarely satisfied in the gas in our simulations. For example, in solar
metallicity gas with $n = 300 \: {\rm cm^{-3}}$, it is satisfied only when the ratio of atomic to ionized
carbon is of the order of a thousand to one. Therefore, in general, the rate at which CH$_{2}^{+}$
ions are produced in the more complex models is slower than the rate at which they are produced
in the simpler models.

This fact explains most of the difference that we see between our two sets of chemical models.
When $n_{\rm C^{+}} \gg n_{\rm C}$, all of the models produce CH$_{2}^{+}$ ions (and hence also
CO molecules) at a very similar rate.  In models G10g, G10ng and NL99, however, the physical conditions  
that allow large CO fractions to be produced (high density and high visual extinction) also allow 
the recombination of C$^{+}$ to be far more effective than the photoionization of atomic carbon, 
meaning that the equilibrium ratio of C to C$^{+}$ in these regions strongly favours atomic carbon.
As the carbon recombines (which occurs on a timescale that is short compared to the CO formation
timescale), the CO formation rate decreases for the reasons outlined above. The same effect does not
occur in the NL97, KC08e or KC08n models because none of these models include the effects of
C$^{+}$ recombination: model NL97 ignores atomic carbon entirely, while in models KC08e and 
KC08n, it is produced only by the photodissociation of CO. Therefore, the CO formation rate in
models NL97, KC08e or KC08n is in general larger than in models G10g, G10ng and NL99, explaining
the shorter CO formation timescales and larger CO fractions we find in the former models with respect
to the latter.

\section{Conclusions}
\label{concl}

In this paper, we have examined several different approaches to modelling
the chemistry of CO in three-dimensional numerical simulations of turbulent
molecular clouds. The simplest approaches (NL97, KC08n, KC08e) include only
a very small number of reactions and reactants, with the KC08e model making
the additional simplifying assumption of chemical equilibrium. The NL99 adds
a number of additional reactions and reactants, but still remains relatively
simple, owing to its use of artifical species (CH$_{x}$, OH$_{x}$) as 
stand-ins for a much wider range of real molecular ions, radicals and 
molecules. Finally, the most complex models examined here
(G10g and G10ng) add a few more non-equilibrium reactants, a large number 
of additional reactants that are assumed to be in chemical equilibrium, 
and a large number of additional reactions. The complexity of these two
models lies close to the upper limit of what is practical to include in a
three-dimensional simulation at present, but still represents a significant
simplification when compared with state-of-the-art one-zone or one-dimensional
chemical models (see e.g.\ the PDR codes discussed in \citealt{pdr07}). 

As one would expect, the simpler one makes the chemical model, the faster
it becomes to compute the chemical evolution of the gas. Models NL97, KC08e
and KC08n allow the simulations to run roughly an order of magnitude faster
than the reference model, while model NL99 yields a runtime three times 
longer than the simple models, but still three to four times shorter than the 
models based on \citet{glo10}.

Looking in more detail at the results of the simple networks, we find
that they tend to produce more CO than the more complex networks, for
the reasons outlined in the previous section. It is also clear that one area
in which the simpler models fall down is in their treatment of atomic carbon.
This is ignored entirely in the NL97 model, and is significantly underproduced 
in the KC08e and KC08n models, owing to the omission of the effects of C$^{+}$ 
recombination from these models. It should however be noted that the
\citet{kc08} chemical model was
designed for use in a study of the thermal balance of prestellar cores
with mean densities that are more than an order of magnitude larger than
the mean densities considered in our simulations. In these conditions, the 
gas is CO-dominated, atomic carbon is not an important coolant, and the
\citet{kc08} model does a good job of capturing the essential physics of 
the gas. It performs more poorly in our study only because we are applying
it to a range of physical conditions outside of those for which it was designed.

Turning our attention to the more complex networks, we find that models
NL99 and G10ng produce very similar results in all of our comparisons,
and that the differences between the results of these models and that
of model G10g are easily understandable as consequences of the inclusion
of grain-surface recombination of C$^{+}$ in the latter, which significantly 
affects the relative abundances of C and C$^{+}$ produced in the gas. 
We also note that the inclusion of this process leads to a significant
overproduction of atomic carbon and hence yields values for the C/CO
ratio that are significantly higher than those observed in real molecular
clouds. Models NL99 and G10ng, on the other hand, yield values for the
C/CO ratio that appear to be consistent with observations.

Given that model NL99 produces very similar values to model G10ng for
the C and CO abundances, but in only one-third of the computational
time, we suggest that the former is a better choice than the latter if one
is primarily interested in the C$^{+}$ to C to CO transition, or in the
details of the resulting CO distribution. On the other hand, the more complex 
G10ng model is a better choice if one is also interested in the distributions of species 
not tracked in the NL99 model, such as CH, CH$^{+}$, OH, H$_{2}$O or O$_{2}$,
and is willing to pay the additional computational cost for tracking these
species. Furthermore, if all one is interested in is an approximate
picture of which regions are likely to have high CO abundances (as may be
of interest in a large-scale simulation of the Galactic ISM, such as
\citealt{dobbs08}),
then {\em any} of these models are likely to do a reasonable job, as the
differences between the CO distributions that they predict are not enormous
and are likely smaller than the uncertainties introduced by the inevitably 
limited resolution possible within large-scale simulations. For this kind of
modelling, it would therefore make sense to choose one of the simpler, faster
models, such as NL97, KC08e or KC08n.

Finally, we have examined the effects of the choice of chemical network
on the density and temperature distributions in our model clouds. We find
that the influence of the chemistry is surprisingly limited. In both the low
and the high density model clouds, much of the cloud volume is filled with 
warm gas ($T > 30 \: {\rm K}$). Most of the carbon in this warm material 
is in the form of C$^{+}$, and hence C$^{+}$ cooling dominates. All of 
the chemical networks that we examined do a good job of representing this
material. The chemistry of the gas plays a significant role in determining
the temperature only if one looks at colder, denser gas, with much higher
C and CO fractions, but even here, changing the chemistry changes the gas
temperatures by at most a factor of two. These small temperature changes
have little effect on the larger-scale gas dynamics, and so the resulting 
density PDF is the same regardless of which chemical model is adopted.

\section*{Acknowledgements}
The authors would like to thank E.\ Keto and W.\ Langer for useful
discussions on the work presented in this paper. The authors acknowledge 
financial support from the Landesstiftung
Baden-W\"urrtemberg via their program International Collaboration II
(grant P-LS-SPII/18), from the German Bundesministerium f\"ur
Bildung und Forschung via the ASTRONET project STAR FORMAT (grant
05A09VHA), from the DFG under grants no.\  KL1358/4 and KL1358/5,
and from a Frontier grant of Heidelberg University sponsored by the German 
Excellence Initiative. The simulations reported on in this paper were 
primarily performed using the {\em Kolob} cluster at the University of 
Heidelberg, which is funded in part by the DFG via Emmy-Noether grant 
BA 3706. Some additional simulations were performed using the 
{\em Ranger} cluster at the Texas Advanced Computing Center, using
time allocated as part of Teragrid project TG-MCA99S024.

\appendix
\section{Revised thermal model}
Our basic treatment of the thermal evolution of the gas is the same as in \citet{glo10}, and
in the interests of avoiding duplication, we refer the reader interested in the full details of 
our model to that paper. In this Appendix, we restrict ourselves to a discussion of the 
improvements we have made to the basic \citet{glo10} model. To briefly summarize, we 
have made three changes. First, we have modified our treatment of the adiabatic index 
$\gamma$ and the mapping from temperature to internal energy (and vice versa) to 
account for the fact that the internal energy levels of H$_{2}$ are not excited at very low
temperatures. Second, we have relaxed the assumption made in \citet{glo10} and \citet{gm10}
of a constant dust temperature, and now solve self-consistently for $T_{\rm dust}$, 
thereby also improving the accuracy of our treatment of energy transfer from the gas to
the grains. Third, and finally, we have improved our treatment of CO rotational cooling
by extending the tabulated cooling function used in the code to lower temperatures. 
All three changes are discussed in more detail in the sections below.
Together, these changes dramatically improve the accuracy with which we can model 
very cold gas in our simulations, and allow us to remove the artificial temperature floor at 
$10 \: {\rm K}$ that was adopted in our previous studies. We now should be able to 
accurately model the temperature of the gas down to values as low as 5~K, and we find 
in practice that there is very little gas below this temperature in any of our
simulations. 

\subsection{Equation of state}
\label{gamma_section}
The specific internal energy $e$ for an ideal gas with a temperature $T$ and 
a partition function $z$ can be written as \citep{boley07}
\begin{equation}
e = \frac{R}{\mu} T^{2} \frac{\partial \ln z}{\partial T},
\end{equation}
where $R = k / m_{\rm p}$ is the gas constant, $k$ is Boltzmann's constant,
$m_{\rm p}$ is the proton mass and $\mu$ is the mean molecular weight
(in units of the proton mass). Because the gas is ideal, its specific heat capacity
at constant volume, $c_{v}$, can be written as $c_{v} = de/dT$. Many previous
numerical studies of star formation have assumed that $c_{v}$ is independent
of temperature, and hence that $e = c_{v} T$. Unfortunately, this assumption is 
not  true for molecular gas, as \citet{boley07} have recently  pointed out. In 
very cold molecular gas, 
essentially all of the $\mHt$ molecules sit in the ortho or para ground state, and the gas 
behaves as if it were monatomic, with $c_{v} = (3/2) R/\mu$. However, as the temperature 
of the gas is increased, first the rotational and then the vibrational energy levels of 
$\mHt$ become populated, and hence $c_{v}$ changes. For this reason, we do not
assume a constant $c_{v}$ in these simulations. Instead, we have constructed a set
of lookup tables that give $e$ as a function of $T$ and $x_{\mHt}$ (the fractional
abundance of $\mHt$), $T$ as a function of $e$ and  $x_{\mHt}$, and the adiabatic
index $\gamma = c_{p} / c_{v}$ as a function of $e$ (or $T$) and $x_{\mHt}$. 
Whenever we must convert from $e$ to $T$ (or vice versa), or require a value for
$\gamma$, we compute it by interpolating between the values stored in these 
lookup tables. In constructing these tables, we have assumed that the $\mHt$
ortho-to-para ratio has its thermal equilibrium value. 

We have verified that by using this approach, we can reproduce the values 
shown for $\gamma$ and $e/R$ in Figures 1 \& 2 of \citet{boley07}. However, we 
caution that the values for these quantities that are used in our actual simulations 
differ slightly from those in these figures, owing to the fact that we account for the
presence of helium in the gas, whereas \citet{boley07} do not.

\subsection{Dust cooling and the dust temperature}
We assume, as in most studies, that the dust is in thermal equilibrium and solve for the
dust temperature $T_{\rm d}$ by finding the value that satisfies the equation of thermal 
balance for the dust:
\begin{equation}
\Gamma_{\rm ext} - \Lambda_{\rm dust} + \Gamma_{\rm gd} + \Gamma_{\rm H_{2}} = 0.
\end{equation}
Here $\Gamma_{\rm ext}$ is the dust heating rate per unit volume due to absorption of
radiation from the interstellar radiation field, $\Lambda_{\rm dust}$ is the radiative cooling
rate of the dust, $\Gamma_{\rm gd}$ is the net rate at which energy is transferred from
the gas to the dust by collisions, and $\Gamma_{\rm H_{2}}$ is the dust heating rate
per unit volume due to H$_{2}$ formation on the grain surfaces. 

To compute $\Gamma_{\rm ext}$, we follow \citet{gold01} and express it as the product
of a optically thin heating rate, $\Gamma_{\rm ext, 0}$, and a dimensionless factor, 
$\chi$, that represents the attenuation of the interstellar radiation field by dust absorption:
\begin{equation}
\Gamma_{\rm ext} = \chi \Gamma_{\rm ext, 0}.
\end{equation}
The optically thin heating rate is given by
\begin{equation}
\Gamma_{\rm ext, 0} = 4\pi {\cal D} \rho \int_{0}^{\infty} J_{\nu} \kappa_{\rm nu}  \, {\rm d}\nu,
\end{equation}
where ${\cal D}$ is the dust-to-gas ratio, $\rho$ is the gas density, $J_{\nu}$ is
the mean specific intensity of the incident interstellar radiation field (ISRF), 
and $\kappa_{\nu}$ is the dust opacity in units of ${\rm cm^{2}} \: {\rm g^{-1}}$.
For our ISRF, we used the radiation field given in \citet{bl94}, which is a 
combination of data from \citet{mmp83} at short wavelengths and \citet{wr91}
at long wavelengths. For our dust opacities, we use the values given in 
\citet{oh94} for grains that have not coagulated and that are coated with thick 
ice mantles.\footnote{Strictly speaking, we would not expect grains exposed to
the full interstellar radiation field to be ice-coated, as observationally we find
that ices appear only for visual extinctions $A_{\rm V} > 3.3$ \citep{whit88}. In
practice, however, our results for $T_{\rm d}$ are not particularly sensitive to
the particular choice of opacities from \citet{oh94}, and so we can safely ignore
this complication.} \citet{oh94} quote opacities only for wavelengths longer than
$1 \: \mu{\rm m}$, and at shorter wavelengths we use values taken from
\citet{mmp83}. Finally, for ${\cal D}$ we take the standard value for solar 
metallicity gas. With these choices, we find that
\begin{equation}
\Gamma_{\rm ext, 0} = 5.6 \times 10^{-24} n \: {\rm erg} \: {\rm s^{-1}} \: {\rm cm^{-3}},
\end{equation}
where $n$ is the number density of hydrogen nuclei.

\begin{figure}
\centering
\epsfig{figure=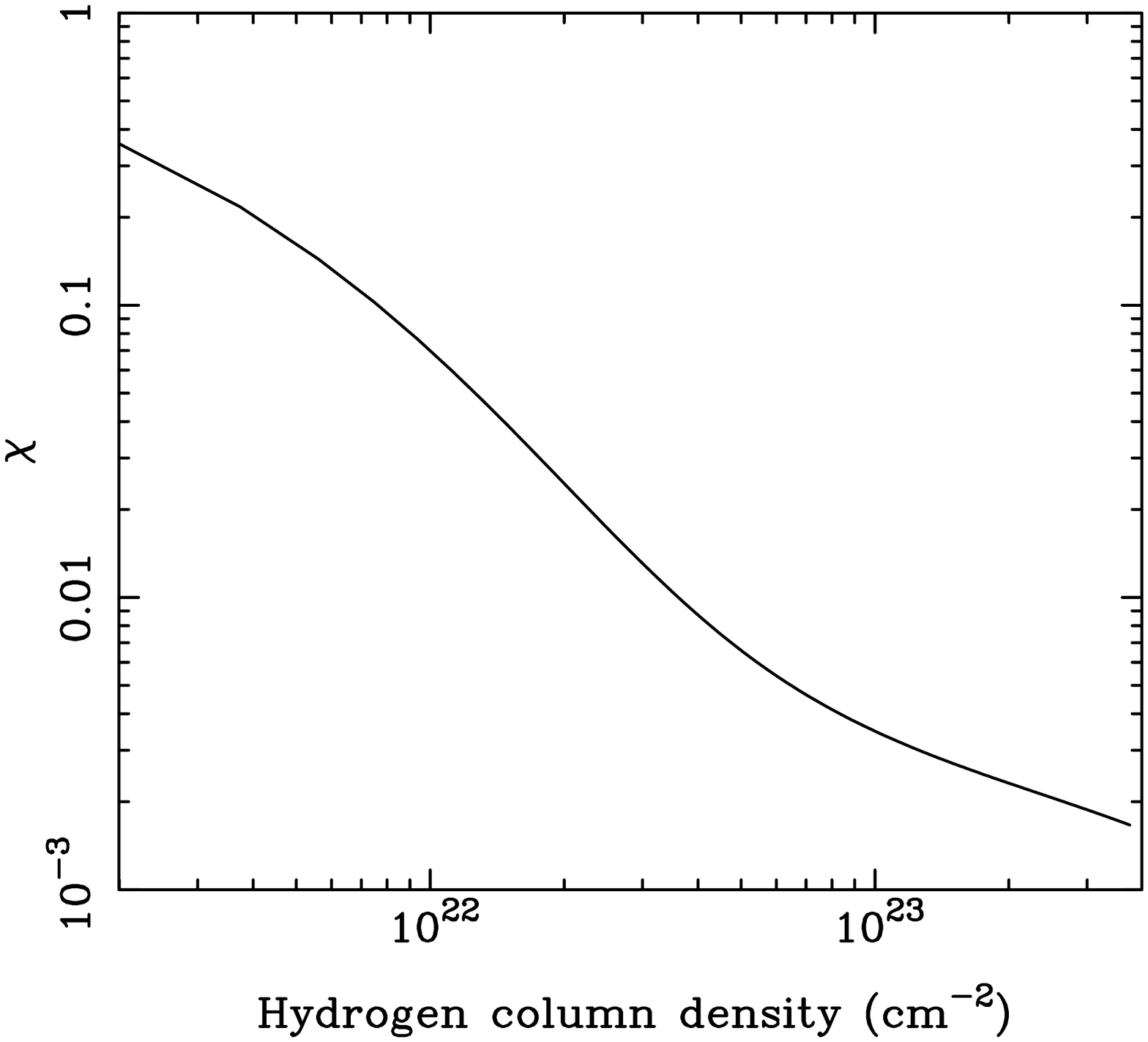,width=15pc,angle=270,clip=}
\caption{Dust attenuation factor $\chi$, plotted as a function of $N_{\rm H}$, the column
density of hydrogen nuclei.  \label{fig:chi}}
\end{figure}

To compute the attenuation factor $\chi$, we solve the equation
\begin{equation}
\chi(N_{\rm H}) = \frac{4\pi \int_{0}^{\infty} J_{\nu} \kappa_{\nu} \exp \left(-\kappa_{\nu} \Sigma \right) 
\, {\rm d}\nu}{4\pi \int_{0}^{\infty} J_{\nu} \kappa_{\nu} \, {\rm d}\nu}, \label{chi_func}
\end{equation}
where $\Sigma = 1.4 m_{\rm p} N_{\rm H}$, $m_{\rm p}$ is the proton mass, and 
$N_{\rm H}$ is the number density of hydrogen nuclei. We solve this equation 
for a wide range of different 
values of $N_{\rm H}$, and tabulate the results. In our simulations, we can then
calculate $\chi$ using our standard six-ray approximation. Recall that this yields,
for each zone in the simulation, the column density of material between the zone
and the edge of the simulation volume along six rays that we take to be parallel
to the coordinate axes, for convenience. We can associate a value of $\chi$ with
each of these column densities, and the appropriate mean value to use in our
calculation of the dust temperature is then simply the arithmetic mean of these
values. By using an approach based on lookup tables, we avoid having to perform
the frequency integration in Equation~\ref{chi_func} during the simulation itself,
for a great saving in computational time. In Figure~\ref{fig:chi}, we show how $\chi$ 
varies with $N_{\rm H}$ for a wide range of values of the column density.

For the dust cooling rate $\Lambda_{\rm dust}$, we solve
\begin{equation}
\Lambda_{\rm dust}(T_{\rm d}) =  4 \pi {\cal D} \rho \int_{0}^{\infty} B_{\nu}(T_{\rm d}) \kappa_{\nu} 
\, {\rm d}\nu,
\end{equation}
where $B_{\nu}(T_{\rm d})$ is the Planck function for a temperature $T_{\rm d}$, and where
our values for $\kappa_{\nu}$ were the same as we used to  calculate $\Gamma_{\rm ext}$ 
above. We find that the resulting cooling rate is well fit by the function
\begin{equation}
\Lambda_{\rm dust}(T_{\rm d}) = 4.68 \times 10^{-31} T_{\rm d}^{6} n \: {\rm erg} \: {\rm s^{-1}} \: 
{\rm cm^{-3}}
\end{equation}
for dust temperatures $5 < T_{\rm d} < 100 \: {\rm K}$, which more than spans the range of
values considered in this study. We do not account for the absorption by dust of its own
thermal radiation when computing  $\Lambda_{\rm dust}(T_{\rm d})$, as 
 we do not follow the evolution of the gas to high enough densities
for this effect to become important. 

For the net rate of energy transfer from gas to dust, $\Gamma_{\rm gd}$, we adopt the expression
\citep{hm89}
\begin{equation}
\Gamma_{\rm gd} =  3.8 \times 10^{-33} T^{1/2} \alpha (T - T_{\rm d}) n^{2}  \: {\rm erg} \: {\rm s^{-1}} \: {\rm cm^{-3}},
\end{equation}
where 
\begin{equation}
\alpha = 1.0 - 0.8 \exp \left(\frac{-75}{T}\right).
\end{equation}

Finally, for the rate at which dust is heated by the formation of H$_{2}$ on grain surfaces,
$\Gamma_{\rm H_{2}}$, we have the expression
\begin{equation}
\Gamma_{\rm H_{2}} = 7.2 \times 10^{-12} f_{\rm dust} R_{\rm H_{2}} \: {\rm erg} \: 
{\rm s^{-1}} \: {\rm cm^{-3}},
\end{equation}
where $R_{\rm H_{2}}$ is the formation rate of H$_{2}$ per unit volume \citep{hm79},
and $f_{\rm dust}$ is the fraction of the 4.48eV H$_{2}$ binding energy that is absorbed
by the grain.  We assume, following \citet{tu01}, that $f_{\rm dust} = 0.04$.

\subsection{CO cooling}
We have extended the range of temperatures for which we can 
calculate CO cooling rates. In \citet{glo10}, we made use of the CO
cooling function developed by \citet{nk93} and \citet{nlm95} and hence
were limited by the range of their tabulated data, which covers the
temperature range $10 < T < 2000$~K for CO rotational cooling. In our present 
study, we have extended their cooling function down to 5~K, using CO 
de-excitation rates taken from \citet{fl01} and \citet{we06} and made available in 
a convenient form by LAMDA, the Leiden atomic and molecular database 
\citep{lamda}.


\begin{thebibliography}{}

\bibitem[Bensch et~al.(2003)]{ben03}
Bensch, F., Leuenhagen, U., Stutzki, J., Schieder, R.\ 2003, ApJ, 591, 1013

\bibitem[Black(1994)]{bl94}
Black, J.~H. 1994, ASP Conf.\ Ser.\ 58, in The First Symposium on the Infrared Cirrus 
and Diffuse Interstellar Clouds, eds.\ R.~M.~Cutri \& W.~B.~Latter, (San Francisco:ASP), 355

\bibitem[Boley et~al.(2007)]{boley07}
Boley, A.~C., Hartquist, T.~W., Durisen, R.~H., \& Michael, S. 2007, ApJ, 656, L89

\bibitem[Dame et~al.(2001)]{dame01}
Dame, T.~M., Hartmann, D., \& Thaddeus, P. 2001, ApJ, 547, 792

\bibitem[Draine(1978)]{dr78}
Draine, B.~T. 1978, ApJS,  36, 595

\bibitem[Dobbs et~al.(2008)]{dobbs08}
Dobbs, C.~L., Glover, S.~C.~O., Clark, P.~C., \& Klessen, R.~S.\ 2008, MNRAS, 389, 1097

\bibitem[Federrath, Klessen \& Schmidt(2008)]{fks08}
Federrath, C., Klessen, R.~S., \& Schmidt, W. 2008, ApJ, 688, L79

\bibitem[Flower(2001)]{fl01}
Flower, D.~R. 2001, J.\ Phys.\ B, 34, 2731

\bibitem[Glover \& {Mac Low}(2007a)]{gm07a}
Glover, S.~C.~O., \& {Mac Low}, M.-M. 2007a, ApJS, 169, 239

\bibitem[Glover \& {Mac Low}(2007b)]{gm07b}
Glover, S.~C.~O., \& {Mac Low}, M.-M. 2007b, ApJ, 659, 1317

\bibitem[Glover et~al.(2010)]{glo10}
Glover, S.~C.~O., Federrath, C., {Mac Low}, M.-M., \& Klessen, R.~S. 2010, MNRAS, 404, 2

\bibitem[Glover \& {Mac Low}(2011)]{gm10}
Glover, S.~C.~O., \& {Mac Low}, M.-M. 2011, MNRAS, 412, 337

\bibitem[Goldsmith(2001)]{gold01}
Goldsmith, P.~F. 2001, ApJ, 557, 736

\bibitem[Habing(1968)]{habing68}
Habing, H.~J. 1968, Bull. Astron. Inst. Netherlands,  19, 421

\bibitem[Heitsch \& Hartmann(2008)]{hh08}
Heitsch, F., \& Hartmann, L. 2008, ApJ, 689, 290

\bibitem[Heitsch et~al.(2008)]{heit08}
Heitsch, F., Hartmann, L.~W., Slyz, A.~D., Devriendt, J.~E.~G.,  
\& Burkert, A. 2008, ApJ, 674, 316

\bibitem[Hennebelle \& Chabrier(2008)]{hc08}
Hennebelle, P., \& Chabrier, G. 2008, ApJ, 684, 395

\bibitem[Hennebelle \& Chabrier(2009)]{hc09}
Hennebelle, P., \& Chabrier, G. 2009, ApJ, 702, 1428

\bibitem[Hollenbach \& McKee(1979)]{hm79}
Hollenbach, D., \& McKee, C.~F. 1979, ApJS,  41, 555

\bibitem[Hollenbach \& McKee(1989)]{hm89}
Hollenbach, D., \& McKee, C.~F. 1989, ApJ, 342, 306

\bibitem[Ikeda et al.(2002)]{ikeda02}
Ikeda, M., Oka, T., Tatematsu, K., Sekimoto, Y., \& Yamamoto, S.\ 2002, ApJS, 139, 467

\bibitem[Ingalls et~al.(1997)]{ingalls97}
Ingalls, J.~G., Chamberlin, R.~A., Bania, T.~M., Jackson, J.~M., Lane, A.~P., \& 
Stark, A.~A.\ 1997, ApJ, 479, 296

\bibitem[Keto \& Caselli(2008)]{kc08}
Keto, E., \& Caselli, P. 2008,  ApJ, 683, 238

\bibitem[Keto \& Caselli(2010)]{kc10}
Keto, E., \& Caselli, P. 2010, MNRAS, 402, 1625

\bibitem[Krumholz \& McKee(2005)]{km05}
Krumholz, M.~R., \& McKee, C.~F. 2005, ApJ, 630, 250

\bibitem[Lee et~al.(1996)]{lee96}
Lee, H.-H., Herbst, E., {Pineau des For\^ets}, G.,
Roueff, E., \& {Le Bourlot}, J. 1996, A\&A, 311, 690

\bibitem[Leung, Herbst \& Huebner(1984)]{lhh84}
Leung, C.~M., Herbst, E., \& Huebner, W.~F. 1984, ApJS, 56, 231

\bibitem[Liszt(2011)]{liszt11}
Liszt, H.\ 2011, A\&A, 527, A45

\bibitem[{Mac Low} et~al.(1998)]{mkbs98}
{Mac Low}, M.-M., Klessen, R.~S., Burkert, A., \& Smith, M.~D. 1998,
Phys.\ Rev.\ Lett., 80, 2754

\bibitem[{Mac Low} \& Klessen(2004)]{mk04}
{Mac Low}, M.-M., \& Klessen, R.~S. 2004, Rev.\ Mod.\ Phys., 76, 125

\bibitem[Mathis, Mezger \& Panagia(1983)]{mmp83}
Mathis, J.~S., Mezger, P.~G., \& Panagia, N. 1983,  A\&A, 128, 212

\bibitem[Millar et~al.(1988)]{mil88}
Millar, T.~J., Defrees, D.~J., McLean, A.~D., \& Herbst, E. 1988, A\&A, 194, 250

\bibitem[Nelson \& Langer(1997)]{nl97}
Nelson, R.~P., \& Langer, W.~D. 1997, ApJ, 482, 796

\bibitem[Nelson \& Langer(1999)]{nl99}
Nelson, R.~P., \& Langer, W.~D. 1999, ApJ, 524, 923

\bibitem[Neufeld \& Kaufman(1993)]{nk93}
Neufeld, D.~A., \& Kaufman, M.~J. 1993, ApJ, 418, 263

\bibitem[Neufeld, Lepp \& Melnick(1995)]{nlm95}
Neufeld, D.~A., Lepp., S., \& Melnick, G.~J. 1995, ApJS, 100, 132

\bibitem[Ossenkopf \& Henning(1994)]{oh94}
Ossenkopf, V., \& Henning, Th. 1994, A\&A, 291, 943

\bibitem[Ostriker, Stone \& Gammie(2001)]{osg01}
Ostriker, E.~C., Stone, J.~M., \& Gammie, C.~F. 2001, ApJ, 546, 980

\bibitem[Padoan, Nordlund \& Jones(1997)]{pnj97}
Padoan, P., Nordlund, A., \& Jones, B.~J.~T. 1997, MNRAS, 288, 145

\bibitem[Padoan \& Nordlund(2002)]{pn02}
Padoan, P., \& Nordlund, A. 2002, ApJ, 576, 870

\bibitem[Padoan \& Nordlund(2011)]{pn09}
Padoan, P., \& Nordlund, A. 2011, ApJ, 730, 40

\bibitem[Pineda, Caselli \& Goodman(2008)]{pineda08}
Pineda, J.~E., Caselli, P., \& Goodman, A.~A. 2008, ApJ, 679, 481

\bibitem[R\"{o}llig et~al.(2007)]{pdr07}
R\"{o}llig, M., et~al., 2007, A\&A, 467, 187

\bibitem[Sch\"oier et~al.(2005)]{lamda}
Sch\"oier, F.~L., {van der Tak}, F.~F.~S., {van Dishoeck}, E.~F. \& Black, J.~H. 2005, A\&A,
432, 369

\bibitem[Sheffer et~al.(2007)]{shef07}
Sheffer, Y., Rogers, M., Federman, S.~R., Lambert, D.~L., \& Gredel, R. 2007, ApJ, 667, 1002

\bibitem[Shetty et~al.(2011)]{shetty10}
Shetty, R., Glover, S.~C.~O., Dullemond, C., \& Klessen, R.~S., 2011, MNRAS, 412, 1686

\bibitem[Takahashi \& Uehara(2001)]{tu01}
Takahashi, J., \& Uehara, H. 2001, ApJ, 561, 843

\bibitem[Tielens \& Hollenbach(1985)]{th85}
Tielens, A.~G.~G.~M., \& Hollenbach, D. 1985, ApJ, 291, 722

\bibitem[Tielens(2005)]{tielens05}
Tielens, A.~G.~G.~M., 2005, The Physics and Chemistry of the Interstellar Medium, (Cambridge: Cambridge Univ. Press)

\bibitem[{van Dishoeck}(1988)]{vd88}
{van Dishoeck}, E.~F. 1988, in `Rate Coefficients in Astrochemistry',
eds.\ T.~J. Millar \& D.~A. Williams, (Kluwer:Dordrecht), 49

\bibitem[{van Dishoeck} \& Black(1988)]{vdb88}
{van Dishoeck}, E.~F., \& Black, J.~F. 1988, ApJ, 334, 771

\bibitem[Wakelam, Herbst \& Selsis(2006)]{wak06}
Wakelam, V., Herbst, E., \& Selsis, F. 2006, A\&A, 451, 551

\bibitem[Wakelam et~al.(2010)]{wak10}
Wakelam, V., Herbst, E., {Le Bourlot}, J., Hersant, F., Selsis, F., \& Guilloteau, S. 2010, A\&A, 517, A21

\bibitem[Weingartner \& Draine(2001)]{wd01}
Weingartner, J.~C., \& Draine, B.~T. 2001, ApJ, 563, 842

\bibitem[Wernli et~al.(2006)]{we06}
Wernli, M., Valiron, P., Faure, A., Wiesenfeld, L., Jankowski, P., \& Szalewicz, K. 2006, A\&A, 446, 367

\bibitem[Whittet et~al.(1988)]{whit88}
Whittet, D.~C.~B., Bode, M.~F., Longmore, A.~J., Adamson, A.~J., McFadzean, A.~D., Aitken, D.~K.,
Roche, P.~F. 1988, MNRAS, 233, 321

\bibitem[Wolfire et~al.(1995)]{wolf95}
Wolfire, M.~G., Hollenbach, D., McKee, C.~F., Tielens, A.~G.~G.~M., \& Bakes, E.~L.~O. 1995,
ApJ, 443, 152

\bibitem[Woodall et~al.(2007)]{umist07}
Woodall, J., Ag\'undez, M., Markwick-Kemper, A.~J., \& Millar, T.~J., 2007, A\&A, 466, 1197

\bibitem[Wright et~al.(1991)]{wr91}
Wright, E.~L., et~al. 1991, ApJ, 381, 200

\end{thebibliography}
\end{document}